\documentclass[useAMS,usenatbib]{mn2e}
\usepackage{graphicx,txfonts,amssymb,amsfonts,natbib,times,hyperref}

\begin{document}

\title[Clustering with non-Gaussian initial conditions]
  {The clustering of galaxy clusters in cosmological models with non-Gaussian initial conditions: Predictions for future surveys}

\author[C.\,Fedeli et al.]{C.\,
      Fedeli$^{1,2,3}$, L.\,Moscardini$^{1,2,3}$ and S.\,Matarrese$^{4,5}$\\$^1$ 
      Dipartimento di Astronomia, Universit\`a di Bologna,
      Via Ranzani 1, I-40127 Bologna, Italy (cosimo.fedeli@unibo.it)\\$^2$ INAF-Osservatorio
      Astronomico di Bologna, Via Ranzani 1, I-40127 Bologna, Italy\\$^3$
      INFN, Sezione di Bologna, Viale Berti Pichat 6/2, I-40127 Bologna, Italy\\$^4$
      Dipartimento di Fisica "G. Galilei", Universit\`a di Padova, Via Marzolo 8, I-35131 Padova, Italy\\$^5$
      INFN, Sezione di Padova, Via Marzolo 8, I-35131 Padova, Italy}

\maketitle

\begin{abstract}
We predict the biasing and clustering properties of galaxy clusters that are expected to be observed in the catalogues produced by two forthcoming X-ray and Sunyaev-Zel'dovich effect surveys. We study a set of flat cosmological models where the primordial density probability distribution shows deviations from Gaussianity in agreement with current observational bounds form the background radiation. We consider both local and equilateral shapes for the primordial bispectrum in non-Gaussian models. The two catalogues investigated are those produced by the \emph{e}ROSITA wide survey and from a survey based on South Pole Telescope observations. It turns out that both the bias and observed power spectrum of galaxy clusters are severely affected in non-Gaussian models with local shape of the primordial bispectrum, especially at large scales. On the other hand, models with equilateral shape of the primordial bispectrum show only a mild effect at all scales, that is difficult to be detected with clustering observations. Between the two catalogues, the one performing better is the \emph{e}ROSITA one, since it contains only the largest masses, that are more sensitive to primordial non-Gaussianity.
\end{abstract}


\section{Introduction}
According to the concordance scenario of structure formation, the virialized cosmic structures that we observe today are the result of gravitationally-induced growth of small, primordial density fluctuations. The mean amplitude of these seed fluctuations can be derived by the level of Cosmic Microwave Background (CMB) temperature anisotropies, and makes up one of the cases for the existence of collisionless dark matter. The question of how these primordial fluctuations are generated is generically answered by assuming a phase of accelerated (inflationary) expansion of the Universe right after the Big Bang, that amplified quantum fluctuations of the matter field to produce the seed fluctuations.

Many different models of inflation have been proposed over the years \citep{GU81.1,LI83.1}, differing mainly in the number and properties of the scalar field(s) driving the accelerated expansion. A generic prediction of many of these models is that the power spectrum of the primordial density fluctuations should be scale-free, with a spectral index very close to the Harrison-Zel'dovich value of unity. Other models predict instead a spectral index dependence on the scale (running index). Also, not all models of inflation predict that primordial density fluctuations are Gaussianly distributed, forecasting instead a probability distribution that is not Gaussian \citep{BA04.1,LO08.1}. CMB constraints on the level of non-Gaussianity do not yet rule out this possibility, hence it is natural to search for alternative probes of it and to explore the observable consequences that deviations from Gaussianity can have on the formation of cosmic structures. Additionally, it is possible for the non-Gaussian amplitude to depend on scale, hence exploring cluster and galaxy scales, that are very different from CMB ones, can help constraining this dependence.

Generically, it is expected that a probability distribution for density fluctuations that is positively skewed would produce a larger number of high density peaks, hence leading to the formation of more numerous massive structures. Similarly, a negatively skewed distribution would have the opposite effect. The consequences of this on the mass function of galaxy clusters, that are naturally the most affected structures, can be substantial, and have been studied in a series of works \citep{MA00.2,VE00.1,MA04.1,KA07.1,GR07.1,MA09.1,GR09.1}. Another, more recently studied effect of primordial non-Gaussianity on non-linear structures is on their spatial distribution. It is expected that in primordial density fluctuations fields with probability density different from Gaussian the density peaks are differently clustered together, leading to different biases and correlation functions of galaxies and galaxy clusters. Only recently has a coherent framework for computing the correction to linear bias in generic models with primordial non-Gaussianity been developed \citep{MA08.1}.

In the present paper we employ these results on cluster biasing and earlier results on the cluster mass function suitably calibrated against fully numerical \emph{n}-body simulations in order to compute the clustering properties of galaxy clusters as would be measured in the cluster catalogues produced by two forthcoming surveys with various models and levels of non-Gaussianity. The cluster catalogues that we adopted are two amongst those that have been addressed recently by \cite{FE08.2}, and they are produced by \emph{e}ROSITA and the South Pole Telescope (SPT henceforth). The former is a space-based X-ray observatory, while the latter is a telescope aimed at millimetric and sub-millimetric observations of the \cite{SU72.1} (SZ) thermal distortion of the CMB spectrum. Both are expected to represent the state of the art instrumentation in the respective fields for the forthcoming decade.

The rest of this work is organized as follows. In Section \ref{sct:ng} we detail the models of primordial non-Gaussianity adopted, including the way in which we computed the mass function and the linear bias of galaxy clusters in those models. In Section \ref{sct:cat} we briefly describe the two cluster catalogues investigated in this paper, specifying the parameters of the related surveys and the scaling relations between cluster mass and X-ray/SZ observables that have been employed. In Section \ref{sct:res} we present our main results, showing the effective bias, observed power spectrum and spatial correlation functions for the non-Gaussian models and comparing them with the Gaussian case. Finally, in Section \ref{sct:con} we summarize our conclusions. The reference Gaussian model we considered is a standard $\Lambda$CDM cosmology with best fit parameters taken from the $5$-years WMAP data release, in conjunction with type Ia supernovae and Baryon Acoustic Oscillation datasets \citep{DU09.1,KO09.1}. The present values of the density parameters for matter, dark energy and baryons are $\Omega_{\mathrm{m},0} = 0.279$, $\Omega_{\Lambda,0} = 0.721$ and $\Omega_{\mathrm{b,0}} = 0.046$ respectively. The Hubble constant reads $H_0 = h 100$ km s$^{-1}$ Mpc$^{-1}$, with $h = 0.701$. The normalization of the cold dark matter power spectrum of primordial density fluctuations is fixed by $\sigma_8 = 0.817$, and the slope thereof is $n = 0.96$.

\section{Primordial Non-Gaussianity}\label{sct:ng}

A particularly simple and useful way to parametrize primordial non-Gaussianity consists in writing the
Bardeen's gauge invariant potential $\Phi$ as the sum of a linear Gaussian term and a non-linear second-order term that encapsulates the deviation from Gaussianity \citep{SA90.1,GA94.1,VE00.1,KO01.1},

\begin{equation}\label{eqn:ng}
\Phi = \Phi_\mathrm{G} + f_\mathrm{NL} * \left( \Phi^2_\mathrm{G} - \langle \Phi_\mathrm{G}^2 \rangle \right).
\end{equation}
In Eq. (\ref{eqn:ng}) the dimensionless parameter $f_\mathrm{NL}$, that weights the quadratic correction to the Gaussian random field $\Phi_\mathrm{G}$, is in general scale and configuration dependent. The symbol $*$ denotes standard convolution, and in the particular case in which $f_\mathrm{NL}$ is constant, it reduces to simple multiplication. Bardeen's potential $\Phi$, on scales smaller than the Hubble radius, equals minus the usual Newtonian gravitational potential.

As recently noted by different authors \citep{AF08.1,PI08.1,CA08.1,GR09.1}, there is some ambiguity in the normalization of Eq. (\ref{eqn:ng}). According to the Large Scale Structure (LSS) convention, that is the one used here, $\Phi$ is linearly extrapolated at $z = 0$. In the CMB convention instead $\Phi$ is primordial, so that $f_\mathrm{NL} = g(+\infty) f_\mathrm{NL}^\mathrm{CMB}/g(0) \simeq 1.3 f_\mathrm{NL}^\mathrm{CMB}$, where $g(z)$ is the linear growth suppression factor for cosmological models different from the Einstein-de Sitter one. It is defined by

\begin{equation}
D_+(z) = \frac{1}{1+z} \frac{g(z)}{g(0)},
\end{equation}
where $D_+(z)$ is the linear growth factor. This means that any constraint on the value of $f_\mathrm{NL}$ gathered from CMB data should be increased by $\sim 30\%$ in order to comply with the convention adopted in this work.

Embracing the nomenclature of \cite{MA08.1}, let us write down the Fourier transform of the present-time linear overdensity filtered on some physical scale $R$.

\begin{equation}\label{eqn:delta}
\delta_R({\bf k}) = \frac{2}{3} \frac{T(k)k^2}{H_0^2 \Omega_{\mathrm{m},0}} W_R(k)\Phi({\bf k}) \equiv \mathcal{M}_R(k) \Phi({\bf k}),
\end{equation}
where $W_R(k)$ is the Fourier transform of the top-hat smoothing function, $T(k)$ is the matter transfer function and $k \equiv \|{\bf k}\|$. For $T(k)$ we adopt the fit of \cite{BA86.1} with the correction due to baryon physics reported in \cite{SU95.1}. More sophisticated fits for the effect of baryons on the matter power spectrum exist \citep{EI98.1}, however we checked that these additional refinements are unimportant at the scales of interest here.

From the above Eq. (\ref{eqn:delta}), it follows that the relation between the power spectrum of matter density fluctuations extrapolated at present, $P({\bf k})$, and the power spectrum of the Newtonian potential, $P_\Phi({\bf k})$, reads

\begin{equation}
P({\bf k}) T^2(k) = \left[ \frac{2}{3} \frac{T(k)k^2}{H_0^2 \Omega_{\mathrm{m},0}} \right]^2 P_\Phi({\bf k}).
\end{equation}
As a consequence, if the primordial matter power spectrum is scale-free, $P({\bf k}) = Ak^{n}$ as in our case, the potential power spectrum can be rewritten as

\begin{equation}
P_\Phi({\bf k}) = \frac{9AH_0^4 \Omega_{\mathrm{m},0}^2}{4} k^{n-4} \equiv B k^{n-4}.
\end{equation}

Several models of inflation predict that the bispectrum of primordial perturbations in the potential assumes a particular shape that is called \emph{local} \citep{LO08.1}, and it is such that the magnitude of the bispectrum itself is maximum when one of the three momenta $({\bf k_1},{\bf k_2},{\bf k_3})$ has a much smaller magnitude than the other two ("squeezed" configuration). In these models $f_\mathrm{NL}$ is a dimensionless constant, and the bispectrum can be written as \citep{CR07.1}

\begin{figure*}
	\includegraphics[width=0.5\hsize]{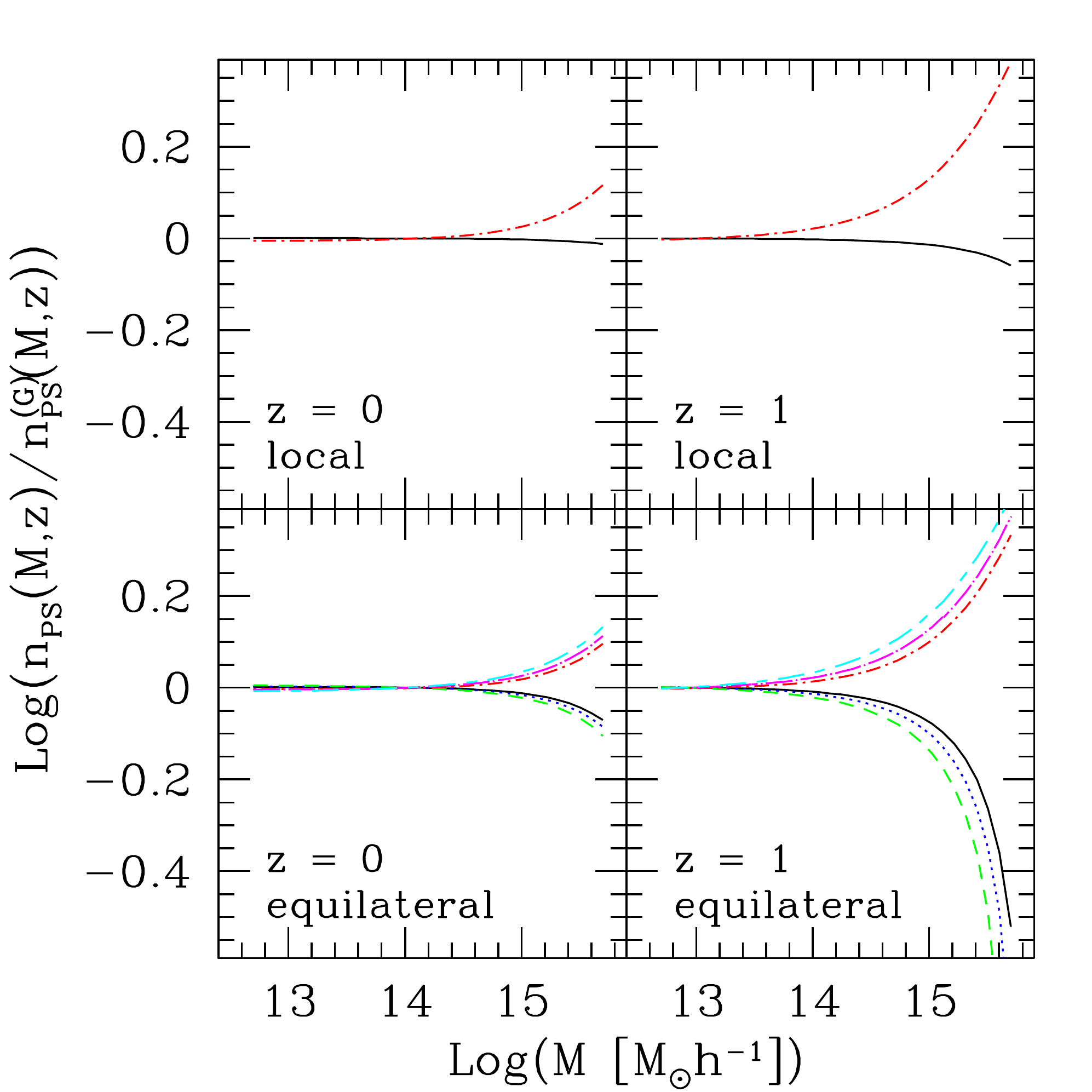}\hfill
	\includegraphics[width=0.5\hsize]{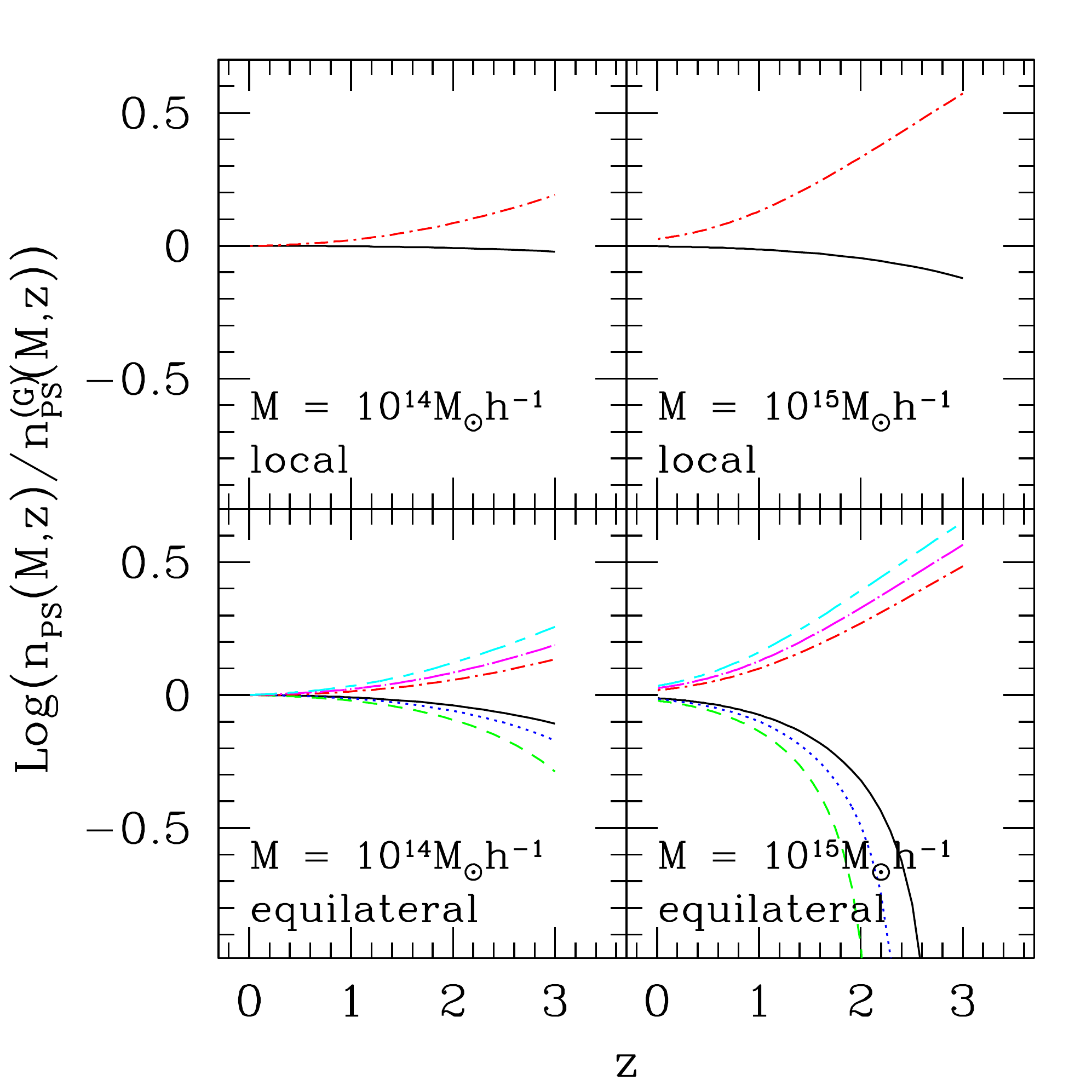}
	\caption{The correction to the mass function for different kinds of non-Gaussian initial conditions. The top four panels refer to a primordial bispectrum with local shape, where the black solid lines have $f_\mathrm{NL} = -12$ and the red dot-dashed curves refer to $f_\mathrm{NL} = 145$. In the four bottom panels we show results for bispectra with equilateral shape, where $f_\mathrm{NL} = -200$ for the bottom group of lines and $f_\mathrm{NL} = 330$ for the topmost group. Different lines refer to different scale dependence for the non-Gaussian amplitude $f_\mathrm{NL}$: $\kappa = 0$ (black solid and red dot-short dashed lines), $\kappa = -0.1$ (blue dotted and magenta dot-long dashed lines) and $\kappa = -0.2$ (green dashed and cyan long-short dashed lines). The four leftmost panels show the amplitude of the correction as a function of mass for two fixed redshifts, as labelled, while the four rightmost ones show the correction as a function of redshift for two fixed masses.}
\label{fig:mfCorrection}
\end{figure*}

\begin{equation}
B_\Phi({\bf k_1},{\bf k_2},{\bf k_3}) = 2f_\mathrm{NL} B^2 \left[ k_1^{n-4}k_2^{n-4} + k_1^{n-4}k_3^{n-4} + k_2^{n-4}k_3^{n-4} \right].
\end{equation}

Other kinds of inflationary scenarios predict a primordial bispectrum with \emph{equilateral shape}, in the sense that it is maximized by configurations where the three arguments have approximately the same magnitude. In the latter case, the primordial bispectrum takes the cumbersome form

\begin{eqnarray}
B_\Phi({\bf k_1},{\bf k_2},{\bf k_3}) &=&  6f_\mathrm{NL} B^2  \left[ k_1^{(n-4)/3}k_2^{2(n-4)/3}k_3^{n-4} \right. + 
\nonumber\\
&+& k_3^{(n-4)/3}k_1^{2(n-4)/3}k_2^{n-4} + k_2^{(n-4)/3}k_3^{2(n-4)/3}k_1^{n-4} + 
\nonumber\\
&+& k_1^{(n-4)/3}k_3^{2(n-4)/3}k_2^{n-4} + k_2^{(n-4)/3}k_1^{2(n-4)/3}k_3^{n-4} + 
\nonumber\\
&+& k_3^{(n-4)/3}k_2^{2(n-4)/3}k_1^{n-4} -k_1^{n-4}k_2^{n-4} - k_1^{n-4}k_3^{n-4} -
\nonumber\\
&-& \left. k_2^{n-4}k_3^{n-4} - 2k_1^{2(n-4)/3}k_2^{2(n-4)/3}k_3^{2(n-4)/3}\right]. 
\end{eqnarray}
Most importantly, in inflationary models that predict an equilateral primordial bispectrum, the parameter $f_\mathrm{NL}$ is in general dependent on the scales. We adopt here the functional form suggested by \cite{LO08.1}, according to which

\begin{equation}\label{eqn:kcmb}
f_\mathrm{NL}({\bf k_1},{\bf k_2},{\bf k_3}) = f_{\mathrm{NL},0} \left( \frac{k_1+k_2+k_3}{k_\mathrm{CMB}} \right)^{-2\kappa}.
\end{equation}
The functional form of Eq. (\ref{eqn:kcmb}) is chosen in order to avoid violating the WMAP constraints, in the sense that $f_{\mathrm{NL},0}$ represents the non-linear parameter evaluated at the scale $k_\mathrm{CMB} = 0.086 h$ Mpc$^{-1}$ roughly corresponding to the largest multipole used by \cite{KO09.1} to estimate non-Gaussianity in the WMAP data, $l = 700$. The constant free parameter $\kappa$ is assumed to be $|\kappa| \ll 1$ between CMB and cluster scales. Consistently with \cite{LO08.1,CR09.1} we assume small and negative values for $\kappa$, that enhance non-Gaussianity on scales smaller than CMB. We adopted $\kappa = 0, -0.1, -0.2$.

\subsection{Mass function}

Generalizations to non-Gaussian models of the standard \cite{PR74.1} mass function have been presented in \cite{MA00.2} and \cite{LO08.1}. Both approaches assume that deviations from Gaussianity are small. In particular, \cite{MA00.2} use the saddle point approximation to compute the probability distribution of threshold crossing, and then truncate the resulting expression to the skewness. \cite{LO08.1} instead approximate the probability density function for the smoothed dark-matter density field using the Edgeworth expansion and then perform the integral of the probability distribution for threshold crossing exactly on the first few terms of the expansion itself. The two approaches give quite similar results, and both have been shown to give reasonable agreement with full numerical simulations of structure formation \citep{GR09.1}, provided the linear overdensity  threshold for collapse is corrected for ellipsoidal density perturbations according to $\Delta_\mathrm{c} \rightarrow \Delta_\mathrm{c} \sqrt{q}$, with $q = 0.75$ (see also \citealt{MA09.1}). In this work, we adopted the formula of \cite{LO08.1}, following which the \cite{PR74.1} mass function for cosmologies with non-Gaussian initial conditions can be written as, setting $\delta_\mathrm{c}(z) \equiv \Delta_\mathrm{c}/D_+(z)$,

\begin{eqnarray}\label{eqn:mfps}
n_\mathrm{PS}(M,z) &=& - \sqrt{\frac{2}{\pi}} \frac{\bar{\rho}(z)}{M} \exp\left[ -\frac{\delta_\mathrm{c}^2(z)}{2\sigma_M^2} \right] \left[ \frac{d\ln \sigma_M}{dM} \left( \frac{\delta_\mathrm{c}(z)}{\sigma_M} + \right.\right.
\nonumber\\
&+& \left. \left. \frac{S_3\sigma_M}{6} \left( \frac{\delta_\mathrm{c}^4(z)}{\sigma^4_M} -2\frac{\delta^2_\mathrm{c}(z)}{\sigma^2_M} -1\right) \right) + \right.
\nonumber\\
&+& \left. \frac{1}{6} \frac{dS_3}{dM}\sigma_M \left( \frac{\delta^2_\mathrm{c}(z)}{\sigma^2_M} -1\right) \right].
\end{eqnarray}

In Eq. (\ref{eqn:mfps}) $\sigma_M$ is the \emph{rms} of the density fluctuations field smoothed on the scale $R$ corresponding to mass $M$, while $S_3(M)$ is the normalized skewness of the same field. It reads $S_3(M) = -f_{\mathrm{NL},0} \mu_3(M)/\sigma_M^4$, where in the local case obviously $f_{\mathrm{NL},0} = f_\mathrm{NL}$, and the third-order moment $\mu_3(M)$ can be computed as

\begin{eqnarray}
\mu_3(M) &=& \int_{\mathbb{R}^9} \mathcal{M}_R(k_1) \mathcal{M}_R(k_2) \mathcal{M}_R(k_3) \times 
\nonumber \\
&\times& B_\Phi({\bf k_1},{\bf k_2},{\bf k_3})\frac{d{\bf k_1}d{\bf k_2}d{\bf k_3}}{(2\pi)^9}.
\end{eqnarray}

Under the assumption that the non-Gaussian correction to the mass function is independent of the approach that is taken to evaluate the mass function itself, the structure abundance in a cosmology with non-Gaussian initial conditions can be computed in compliance to a generic prescription according to

\begin{equation}\label{eqn:mf}
n(M,z) = n^\mathrm{(G)}(M,z) \frac{n_\mathrm{PS}(M,z)}{n_\mathrm{PS}^\mathrm{(G)}(M,z)}.
\end{equation}
In the previous Eq. (\ref{eqn:mf}),  $n_\mathrm{PS}^\mathrm{(G)}(M,z)$ is the mass function computed according to the \cite{PR74.1} formula, while $n^\mathrm{(G)}(M,z)$ is the one computed as specified by the preferred prescription, that in our case was the one detailed in \cite{SH02.1}, both of them evaluated within the Gaussian model. The \cite{PR74.1} mass function in the non-Gaussian cosmologies, $n_\mathrm{PS}(M,z)$, can be computed analytically following Eq. (\ref{eqn:mfps}).

\begin{figure*}
	\includegraphics[width=0.5\hsize]{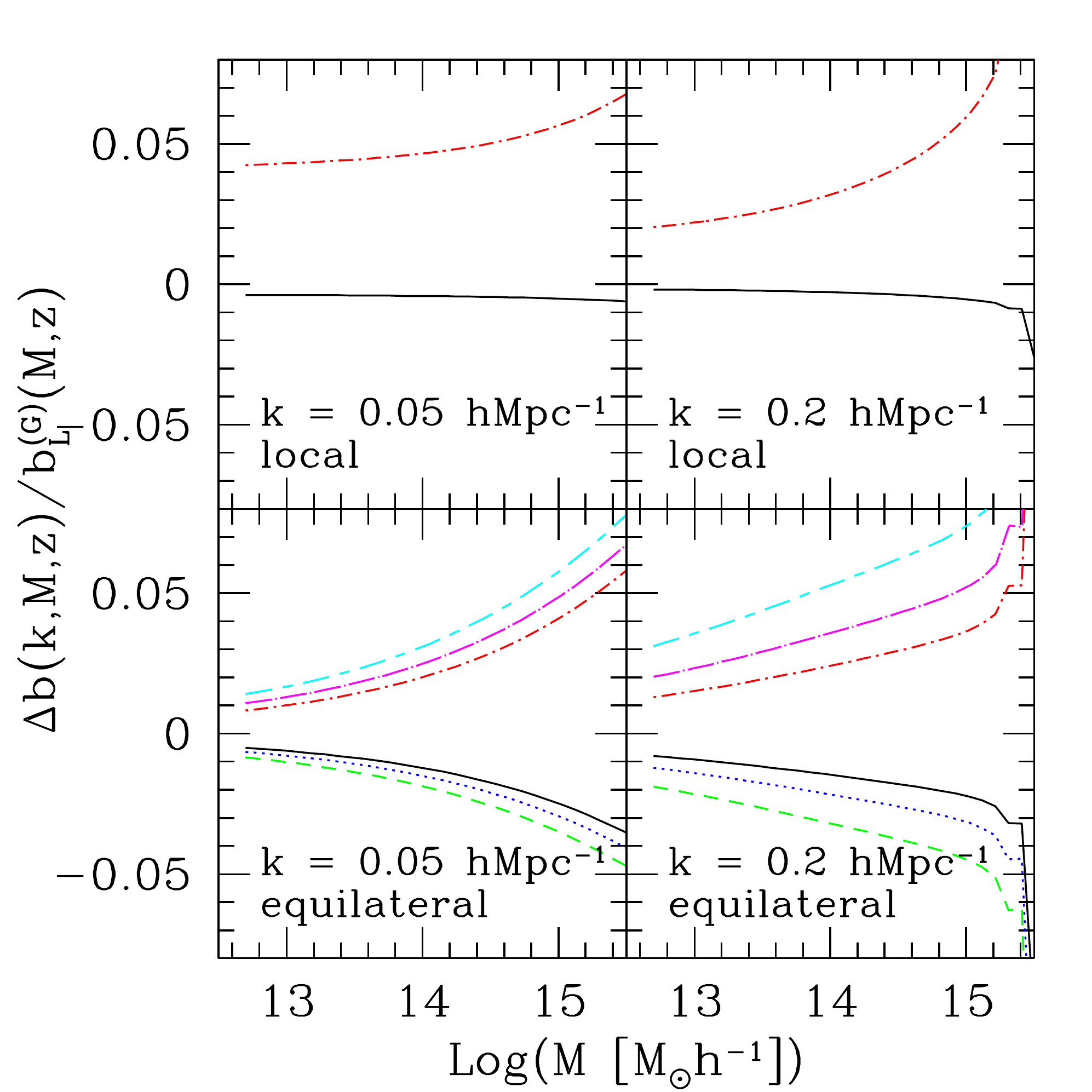}\hfill
	\includegraphics[width=0.5\hsize]{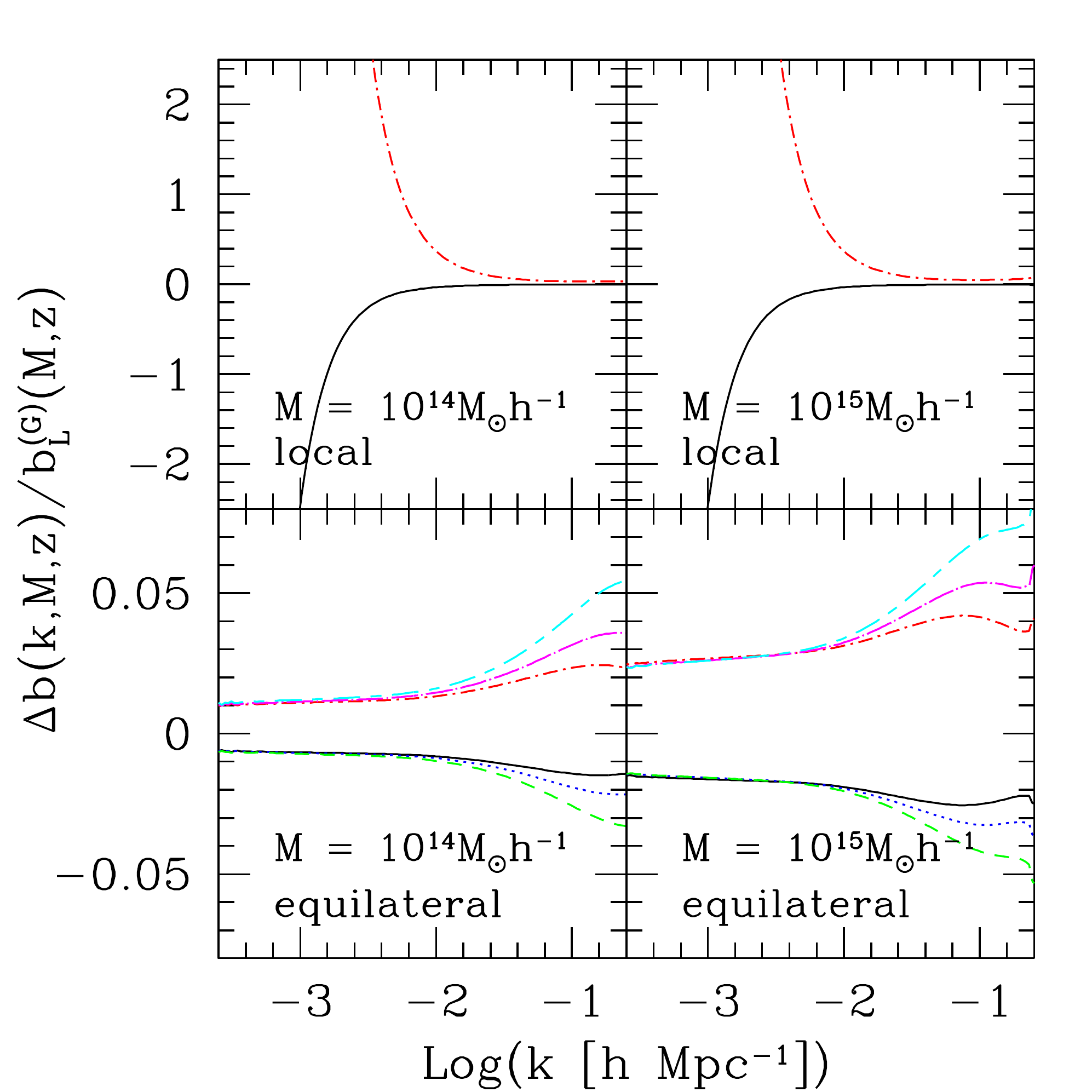}
	\caption{The scale dependent correction to the linear bias for different kinds of non-Gaussian initial conditions. The top four panels refer to a primordial bispectrum with local shape, where the black solid lines have $f_\mathrm{NL} = -12$ and the red dot-dashed curves refer to $f_\mathrm{NL} = 145$. In the four bottom panels we show results for bispectra with equilateral shape, where $f_\mathrm{NL} = -200$ for the bottom group of lines and $f_\mathrm{NL} = 330$ for the topmost group. Different lines refer to different scale dependence for the non-Gaussian amplitude: $\kappa = 0$ (black solid and red dot-short dashed lines), $\kappa = -0.1$ (blue dotted and magenta dot-long dashed lines) and $\kappa = -0.2$ (green dashed and cyan short-long dashed lines). The four leftmost panels show the amplitude of the correction as a function of mass for two fixed scales, as labelled, while the four rightmost ones show the correction as a function of scale for two fixed masses. The redshift is always set to $z = 0$, since it can be easily scaled.}
\label{fig:bsCorrection}
\end{figure*}

In Figure \ref{fig:mfCorrection} we report the corrections to be applied to the Gaussian mass function in order to obtain the non-Gaussian one. As mentioned above, in the equilateral case we adopt three different values for the exponent $\kappa$, namely $\kappa = 0, -0.1$ and $-0.2$. As for the values of $f_\mathrm{NL}$ in the local model and $f_{\mathrm{NL},0}$ in the equilateral models, we adopt the constraints given by the $5$-years WMAP dataset \citep{KO09.1}, that are the tightest presently available, with the exception of the work of \cite{SM09.1} that however quote limits only for the local shape. According to \cite{KO09.1}, for local non-Gaussianity $f_\mathrm{NL}^\mathrm{CMB}$ is allowed to vary between $-9$ and $111$ on CMB scale. Since we are adopting the LSS convention, we used $f_\mathrm{NL} = -12$ and $f_\mathrm{NL} = 145$ as extremal values. Similarly, for the equilateral cases the CMB constraints give $-151 \le f_{\mathrm{NL},0}^\mathrm{CMB} \le 253$, hence we adopted the extremal values $f_{\mathrm{NL},0} = -200$ and $f_{\mathrm{NL},0} = 330$.

Different theoretical studies, based both on analytic and numerical investigations, have addressed the capability of different observables in constraining the non-Gaussian amplitude, such as the abundance of massive virialized structures \citep{MA00.2,VE00.1,MA04.1,KA07.1,GR07.1}, halo biasing (\citealt{DA08.1,MC08.1} and this work), galaxy bispectrum \citep{SE07.2,JE09.1}, mass density distribution \citep{GR08.2} and topology \citep{MA03.2,HI08.2}, integrated Sachs-Wolfe effect \citep{AF08.1,CA08.1}, Ly$\alpha$ flux from low-density intergalactic medium \citep{VI09.1}, $21$-cm fluctuations \citep{CO06.2,PI07.1} and reionization \citep{CR09.1}. In general, all these methods provide weaker constraints than the CMB data, with an interesting exception being \cite{SL08.1}. 

As can be seen in Figure \ref{fig:mfCorrection}, the correction to the mass function increases both with mass and redshift. At $z = 0$ it can be up to $\sim 25\%$ for the most extreme masses, irrespective of the chosen shape for the primordial bispectrum. At $z = 1$ the corrections raise up to a factor of $\sim 2.5$ for the local shape and the equilateral one with positive $f_{\mathrm{NL},0}$. For negative $f_{\mathrm{NL},0}$, the corrections become arbitrarily large for the most extreme masses and high redshifts. This fact is not a concern, since at very high masses and redshifts, the abundance of objects is virtually zero.

\subsection{Bias}

Primordial density fluctuations with non-Gaussian probability distribution cause a scale-dependent modification to the linear bias for given mass and redshift. Hereafter, we adopt the approach detailed in \cite{MA08.1}, according to which we define the function $\mathcal{F}_R(k)$ as

\begin{eqnarray}
\mathcal{F}_R(k) &\equiv& \frac{1}{8\pi^2\sigma_R^2}\int_0^{+\infty}\zeta^2 \mathcal{M}_R(\zeta) \times
\nonumber\\
&\times& \left[ \int_{-1}^{1} \mathcal{M}_R(\sqrt{\alpha}) \frac{B_\Phi(\zeta,\sqrt{\alpha},k)}{P_\Phi(k)}d\mu \right]d\zeta,
\end{eqnarray}
where $\sigma_R$ is the \emph{rms} of density fluctuations filtered on the scale $R$, $\alpha = \zeta^2 + k^2 + 2 \zeta k \mu$ and $B_\Phi$ is the bispectrum of the non-Gaussian potential, where the three arguments have been replaced by scalars, since $B_\Phi({\bf k_1},{\bf k_2},{\bf k_3}) = B_\Phi(k_1,k_2,k_3)$ for both non-Gaussian shapes considered in this work. An important feature to be explored of cosmological models with non-Gaussian initial conditions is the configuration dependence of higher order correlation functions. 

Given all the above, the Eulerian bias in models with non-Gaussian initial conditions can be written as
\begin{equation}\label{eqn:bs}
b(M,z,k) = 1 + b^{(\mathrm{G})}_\mathrm{L}(M,z) \left[ 1 + \frac{\Delta b(M,z,k)}{b^{(\mathrm{G})}_\mathrm{L}(M,z)} \right],
\end{equation}
where the Lagrangian bias in the Gaussian model was assumed to take the form

\begin{eqnarray}
b^{(\mathrm{G})}_\mathrm{L}(M,z) &=& b^{(\mathrm{G})}(M,z) - 1 = a\frac{\Delta_\mathrm{c}}{D_+^2(z)\sigma^2_M} - \frac{1}{\Delta_\mathrm{c}} +
\nonumber\\
&+& \frac{2p}{\Delta_\mathrm{c}} \left[ \frac{[D_+(z)\sigma_M]^{2p}}{[D_+(z)\sigma_M]^{2p} + [\sqrt{a} \Delta_\mathrm{c}]^{2p}} \right],
\end{eqnarray}
(see \citealt{MO96.1,SH99.1,SH01.1}). The parameters are here set to $p = 0.3$ and $a = q = 0.75$. The correction inside the square brackets in Eq. (\ref{eqn:bs}) is

\begin{equation}\label{eqn:bsc}
 \frac{\Delta b(M,z,k)}{b^{(\mathrm{G})}_\mathrm{L}(M,z)} = \frac{\Delta_c}{D_+(z)} \frac{\mathcal{F}_R(k)}{\mathcal{M}_R(k)}.
\end{equation}

In the particular case of a local primordial bispectrum, the relation for $\mathcal{F}_R(k)$ can be simplified to

\begin{eqnarray}
\mathcal{F}_R(k) &=& \frac{2f_\mathrm{NL}}{8\pi^2\sigma_R^2}\int_0^{+\infty}\zeta^2 \mathcal{M}_R(\zeta) P_\Phi(\zeta )\times
\nonumber\\
&\times& \left[ \int_{-1}^{1} \mathcal{M}_R(\sqrt{\alpha}) \left( \frac{P_\Phi(\sqrt{\alpha})}{P_\Phi(k)}  + 2\right)d\mu \right]d\zeta.
\end{eqnarray}

This prescription for the correction to the linear bias has been confronted with \emph{n}-body numerical simulations in \cite{DE09.1}, where it was found a disagreement between the theory and the numerical experiments for some ranges of bias and scale. However, similarly to what happen for the non-Gaussian  mass function, \cite{GR09.1} found that instead a reasonable agreement can be reached with the position $\Delta_\mathrm{c} \rightarrow \Delta_\mathrm{c} q$, obtaining results also in qualitative agreement with \cite{PI08.1}. We adopted this position when computing the bias in non-Gaussian models in this work.

In Figure \ref{fig:bsCorrection} we show the correction to the linear bias as a function of mass and scale. The redshift in this figure is always fixed at $z = 0$, since the correction scales simply as $\delta_\mathrm{c}(z)$. The shape of the correction factor as a function of the scale is in qualitative agreement with the work of \cite{TA08.1}, while a precise quantitative comparison cannot be made due to the different set of parameters that have been used. In particular, we note the expected fact that the correction to the linear bias in the case of local shape grows as $\propto k^{-2}$ at small $k$ \citep{MA08.1}. Conversely, the correction decreases with decreasing scale in the case of equilateral shape. In the latter case, the correction is also much smaller, reaching at most $\sim 10\%$ at small scales and extreme masses. 

\section{Cluster catalogues}\label{sct:cat}

Evaluating the clustering properties of galaxy clusters we referred to two forthcoming surveys, one in the X-ray band and the other one in the millimeter regime, exploiting the thermal SZ distortion. The first one is the wide \emph{e}ROSITA survey, while the second is the SPT survey. These are the two most promising in order to distinguish models with a strong redshift evolution of dark-energy by using the cluster correlation functions, out of the five considered in \cite{FE08.2}. 

The \emph{e}ROSITA wide survey is planned to have a sky coverage of $\sim 2 \times 10^4$ square degrees down to a limiting X-ray flux of $F_\mathrm{lim} = 3.3 \times 10^{-14}$ erg s$^{-1}$ cm$^{-2}$ in the energy band $[0.5,2.0]$ keV (see also the dark-energy task force white paper \citealt{HA05.1}). In order to convert this limiting flux into a minimum mass at fixed redshift we employed the set of scaling relations described in \cite{FE08.2} (see also \citealt{BA03.1,FE07.2}). They consist of the virial relation between mass and X-ray temperature with normalization based on the simulations by \cite{MA01.1}, together with the luminosity-temperature relation required by the observations of \cite{AL98.1}. These imply a relation between mass and bolometric X-ray luminosity of the kind

\begin{equation}\label{eqn:lm}
L(M,z) = 3.087 \times 10^{44} \mathrm{erg s}^{-1} h^{-2} \left[ \frac{M}{10^{15}M_\odot}h(z) \right]^{1.554},
\end{equation}
where the mass is expressed in units of $M_\odot$. In this derivation it is implicitly assumed that the luminosity-temperature relation does not evolve with redshift, as justified by the studies of \cite{MU97.1,RE99.1,HA02.1}, and that the steepening of this relation at galaxy group scales does not apply \citep{OS04.1,KH07.1}.

\begin{figure*}
	\includegraphics[width=0.45\hsize]{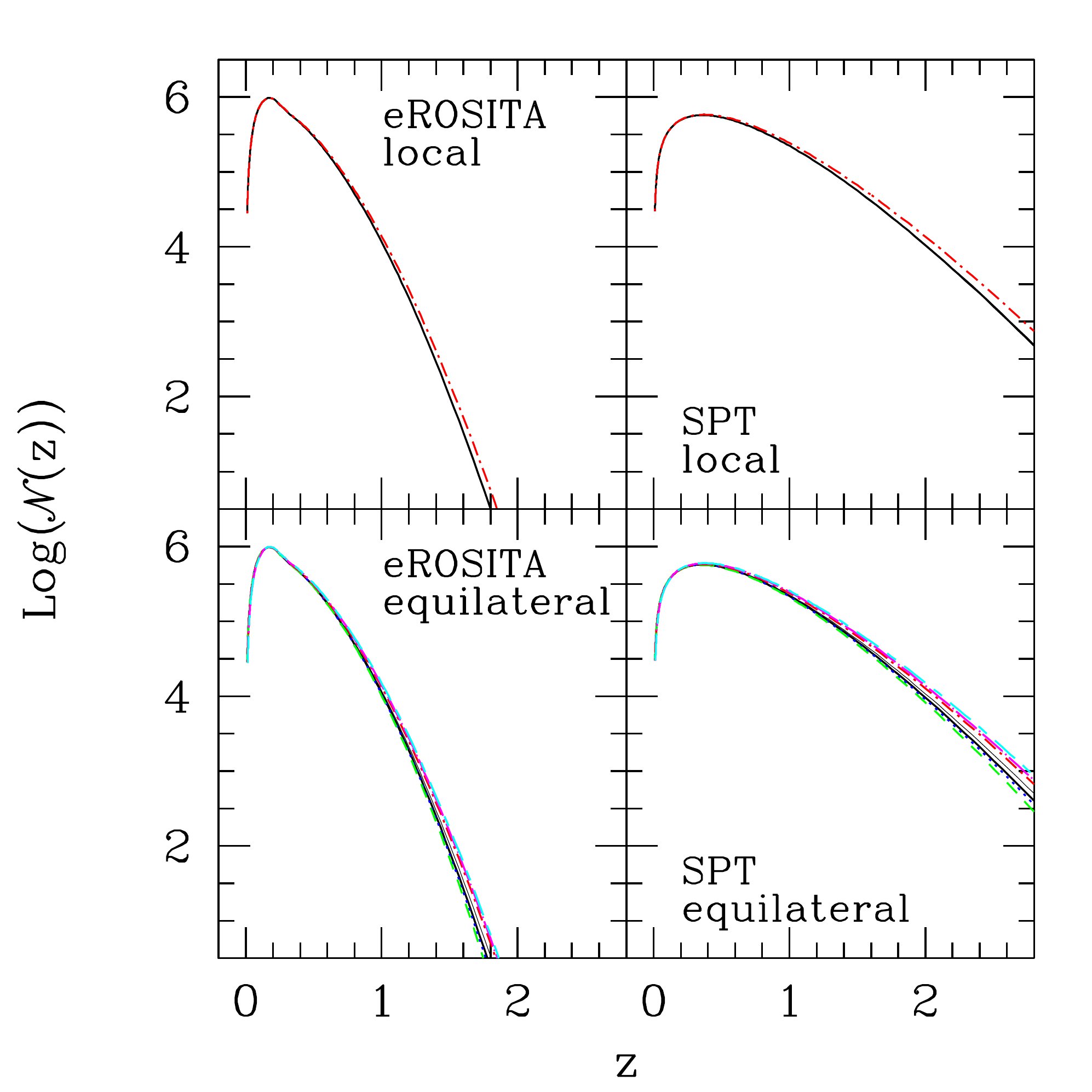}
	\includegraphics[width=0.45\hsize]{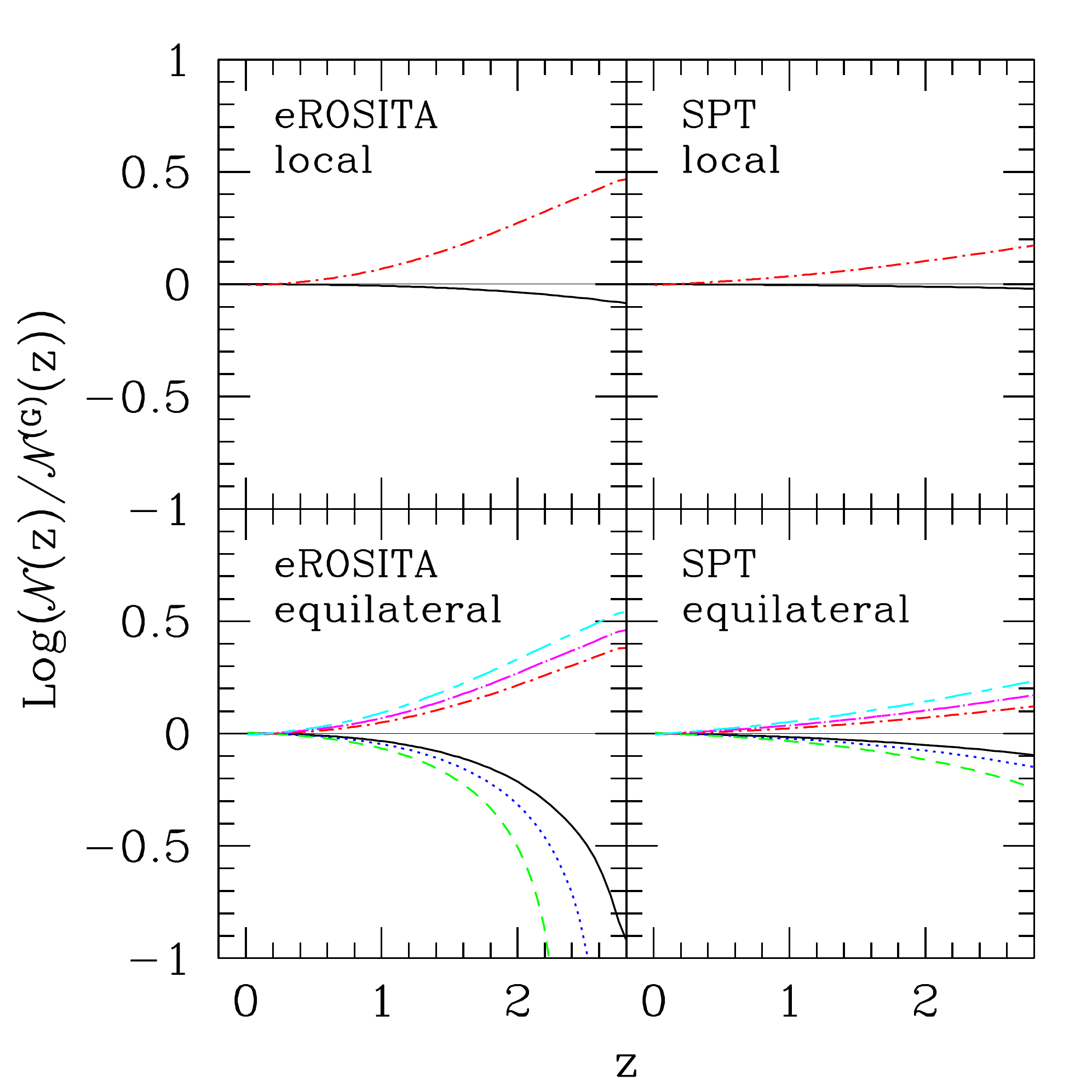}\hfill
	\caption{\emph{Left panel}. The all-sky equivalent redshift distribution of galaxy clusters in the catalogues produced by the \emph{e}ROSITA and SPT surveys. The Gaussian model is represented by the thin black line. \emph{Right panel}. The ratio between the redshift distributions of clusters inside \emph{e}ROSITA and SPT catalogues obtained in different models with non-Gaussian initial conditions to the Gaussian case (thin black line). Line types and colors are the same as in previous Figures.}
\label{fig:rd}
\end{figure*}

We converted the bolometric luminosity implied by Eq. (\ref{eqn:lm}) into a band luminosity by modeling the intra-cluster plasma with a \cite{RA77.1} model with a metal abundance $Z = 0.3 Z_\odot$ \citep{FU98.1,SC99.1}. The plasma model has been implemented with the \texttt{xspec} software package \citep{AR96.1}. The luminosity is then converted into a flux by using the luminosity distance in the appropriate cosmology.

For the SPT catalogue the predicted survey area is of $\sim 4 \times 10^3$ square degrees and we adopted the specifics detailed by \cite{MA03.1}, according to which the limiting SZ flux density at a frequency $\nu_0 = 150$ GHz is $S_{\nu_0,\mathrm{lim}} = 5$ mJy. It is likely that eventually the survey area will be larger than predicted in \cite{MA03.1} (M. Joy, private communication, see also \citealt{ST08.1}), however this is relevant only for evaluating the uncertainties on the observed correlation functions.

In order to link the minimum SZ flux density with a minimum catalogue mass, we used the scaling relation between mass and Compton $y$-parameter integrated over the solid angle covered by the virial sphere of the cluster given by \cite{SE07.1}, together with the relation between the integrated Compton parameter and the nominal SZ flux density. The result is

\begin{equation}
S_{\nu_0}(M,z) = \frac{2.592\times 10^8 \mathrm{mJy}}{(D_\mathrm{A}(z)/1 \mathrm{Mpc})^2} \left( \frac{M}{10^{15}M_\odot} \right)^{1.876} E(z)^{2/3},
\end{equation}
where the mass is again expressed in units of $M_\odot$, $D_\mathrm{A}(z)$ is the angular diameter distance out to redshift $z$ and $E(z) \equiv h(z)/h$. We remark that not all the features of the scaling relations described above are well established, especially concerning their redshift evolution. We however believe that they are the most suitable given our aims.

The minimum mass included in the \emph{e}ROSITA and SPT catalogues as a function of redshift for a variety of dark-energy cosmologies (including standard $\Lambda$CDM model) is shown in \cite{FE08.2}. Since the scaling relations adopted here depend only on the geometry of the Universe, the minimum mass in each of the non-Gaussian models adopted in this work is the same. In Figure \ref{fig:rd} we show the redshift distributions for the \emph{e}ROSITA and SPT catalogues in the various models with non-Gaussian initial conditions, as well as the ratio thereof with respect to the standard $\Lambda$CDM cosmology. The (all-sky equivalent) redshift distribution is defined as

\begin{equation}
\mathcal{N}(z) = 4\pi G(z)\int_{M_1}^{M_2} n(M,z) dM,
\end{equation}
where $n(M,z)$ is the differential mass function and $G(z)$ is the volume contained in the unit redshift, that in a flat Universe can be expressed as the Jacobian determinant

\begin{equation}\label{eqn:jacobian}
G(z) = r^2(z) \frac{dr}{dz}(z).
\end{equation}
In Eq. (\ref{eqn:jacobian}) the function $r(z)$ is the comoving radial distance out to redshift $z$.

The difference between the two catalogues is evident, in that the redshift distribution drops to zero already at $z \lesssim 2$ in the \emph{e}ROSITA catalogue, while it is still significant at $z \sim 3$ in the SPT one. As already discussed by \cite{FE08.2}, this is due to the different redshift dependence of the scaling relations adopted, in particular by the fact that the X-ray flux drops as the square of the luminosity distance, while the SZ flux density drops as the square of the angular-diameter distance.

The difference between different initial condition models are mostly visible in the right panel of Figure \ref{fig:rd}, showing the ratio with respect to the standard $\Lambda$CDM cosmology. In general, for both local and equilateral shapes of the primordial bispectrum, differences are more enhanced in the \emph{e}ROSITA catalogue than in the SPT one, and this is an obvious consequence of the fact that the minimum mass included in the former is larger at any given redshift. Since the deviations from the Gaussian mass function increase with mass (see Figure \ref{fig:mfCorrection}), it is expected that the corresponding redshift distribution is more sensitive.

For the case of local bispectrum with negative $f_\mathrm{NL}$, the deviations with respect to the Gaussian model are always very small, due to the fact that in this case $f_\mathrm{NL}$ is very close to zero. On the other hand, the deviations are more appreciable when $f_\mathrm{NL}$ is positive. For the equilateral shape and $f_\mathrm{NL} < 0$, the departures from the Gaussian cosmology become arbitrarily large in the \emph{e}ROSITA catalogue, however this happens at $z \gtrsim 2$, where the number of objects in the catalogue is practically vanishing. If we limit analysis at $z \lesssim 2$ for the \emph{e}ROSITA catalogue we can see that deviations from the Gaussian redshift distributions are at most of a factor of $\sim 2.5$ in this catalogue. For the SPT catalogue, departures from the Gaussian redshift distribution reach up to $\sim 80\%$ at the highest redshifts, $z \sim 3$, where we still have $\lesssim 10^3$ objects in the catalogue.

\section{Results}\label{sct:res}

In this section we first of all assess the evolution of the effective bias, both as a function of redshift and of scale, in case the initial conditions are not Gaussian. The effective bias is basically given by the linear "monochromatic" bias weighted for the object abundance, and can be written as

\begin{equation}\label{eqn:eb}
b_\mathrm{eff}(z,k) = \frac{4\pi G(z)}{\mathcal{N}(z)} \int_{M_1}^{M_2} b(M,z,k) n(M,z) dM.
\end{equation}
In Eq. (\ref{eqn:eb}), $M_1$ and $M_2$ are the extrema of the mass interval that is encompassed by the catalogue at hand. In the realistic situations we are dealing with, $M_1$ is the minimum mass of a certain catalogue at the given redshift \citep{FE08.2}, while formally $M_2 = +\infty$.

\begin{figure*}
	\includegraphics[width=0.45\hsize]{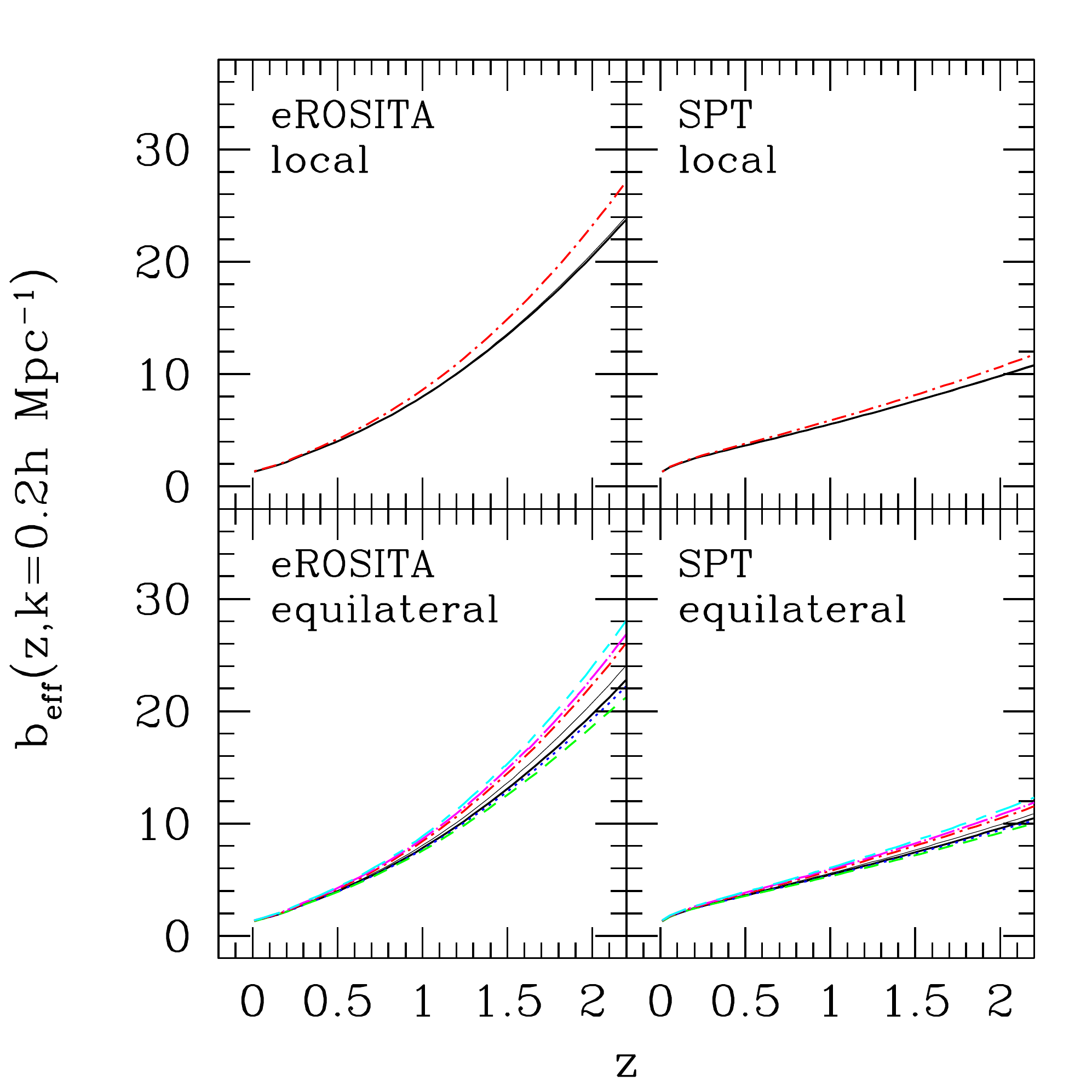}
	\includegraphics[width=0.45\hsize]{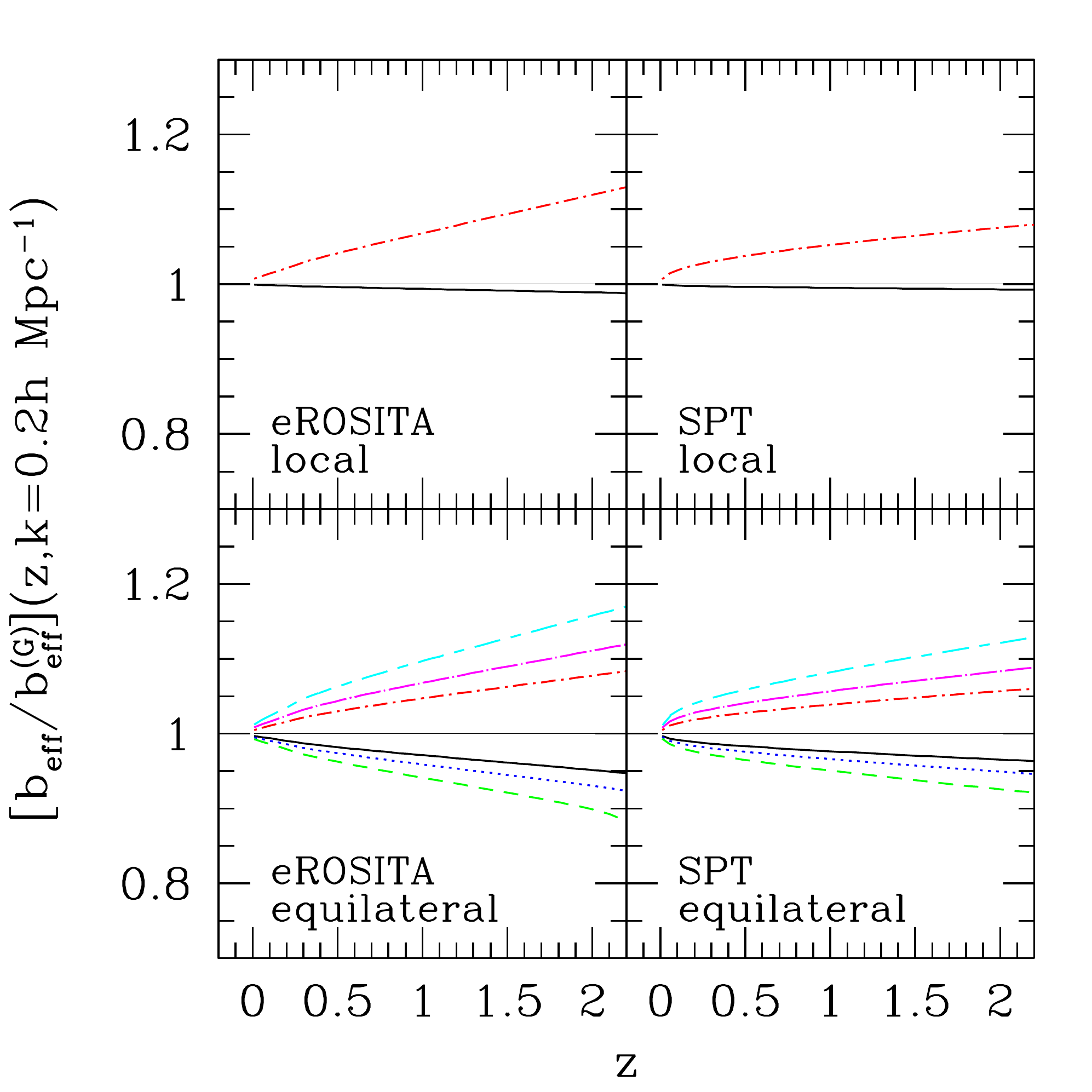}
	\includegraphics[width=0.45\hsize]{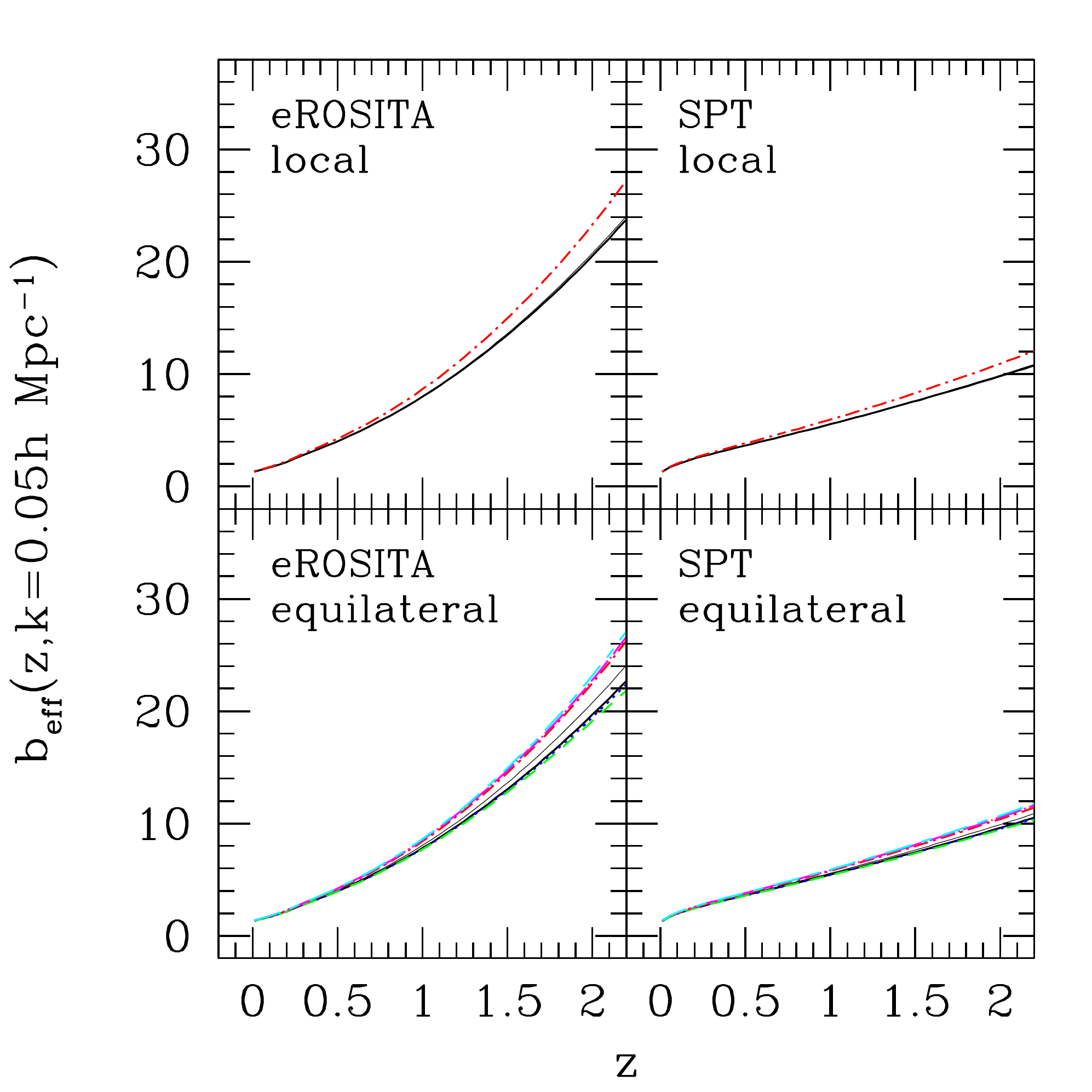}
	\includegraphics[width=0.45\hsize]{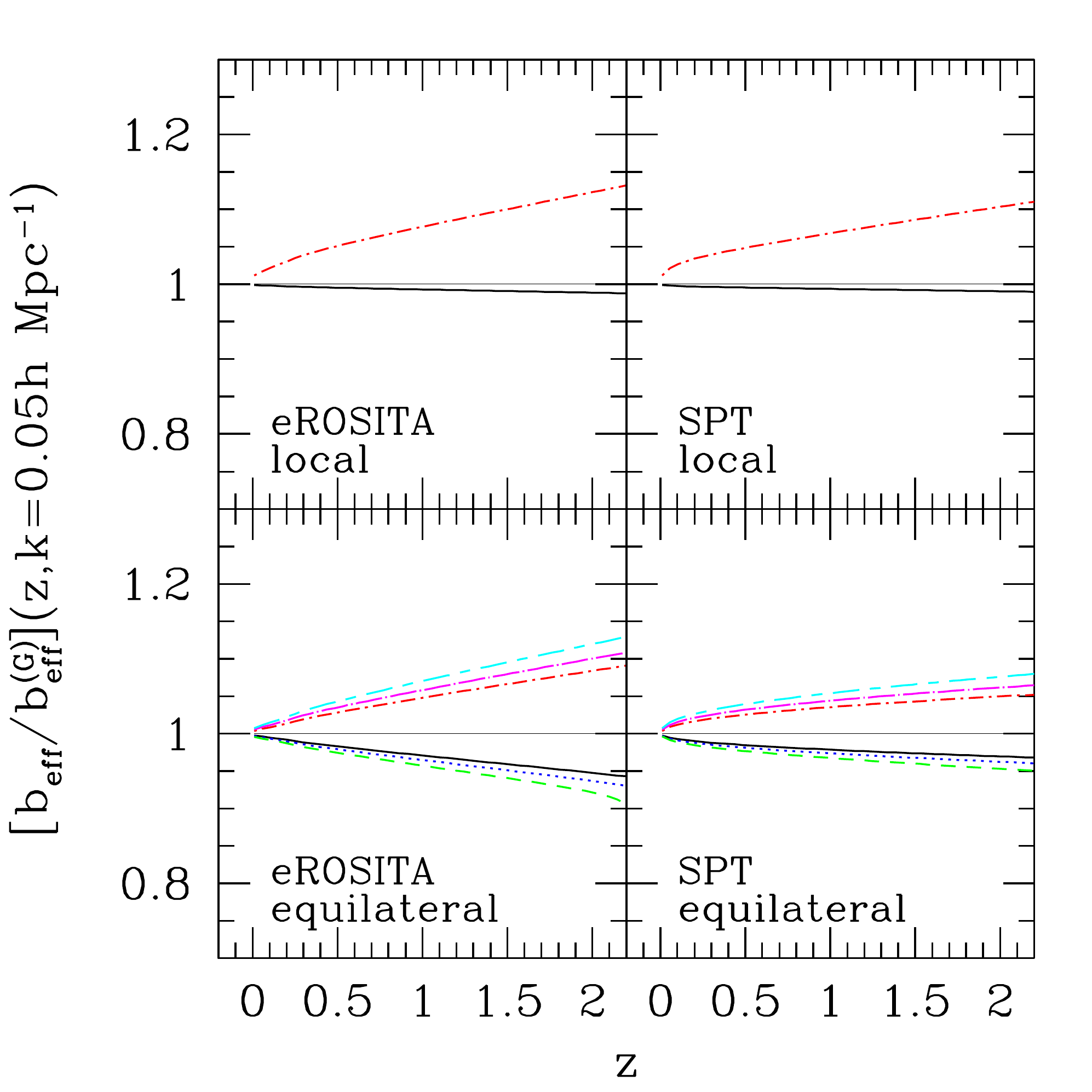}\hfill
	\caption{The effective bias of galaxy clusters included in the catalogues produced with the \emph{e}ROSITA and SPT surveys considered in the present work. The thin black lines refer to the Gaussian model, while the other lines refer to initial conditions of non-Gaussian type, with colors and line types equal to those used in previous Figures. The scale is fixed to $k = 0.2h$ Mpc$^{-1}$ in the top group of panels and to $k = 0.05h$ Mpc$^{-1}$ in the bottom group, as labeled. The four rightmost panels in both rows show the ratio between the non-Gaussian effective biases and the Gaussian one.}
\label{fig:bias_k}
\end{figure*}

The fundamental difference between this formula and the equivalent one for Gaussian initial conditions is that here the effective bias has a dependence on the scale in addition to the redshift dependence. Such scale dependence is inherited by the correction to the monochromatic bias. It is interesting to explore how big this scale dependence is, in order to understand if it could be detectable by looking at the spatial distribution of clusters. In Figure \ref{fig:bias_k} we show the redshift dependence of the effective bias for fixed scales, and the ratio thereof with respect to the Gaussian case. We selected the two fixed scales at $k = 0.05 h$ Mpc$^{-1}$ and $k = 0.2 h$ Mpc$^{-1}$, in order to probe linear and mildly non-linear regimes.

One first thing to note, that might seem counter-intuitive, is that those non-Gaussian models for which the abundance of objects is larger, i.e. those with positive $f_\mathrm{NL}$, are also those where galaxy clusters are more biased with respect to the underlying matter density field. While this is an obvious consequence of the fact that the sign of the correction to the monochromatic bias given in Eq. (\ref{eqn:bsc}) depends on the sign of $f_\mathrm{NL}$, one would naively expect that in models where it is easier for a density peak to collapse into a virialized structure, the structures would be less biased. This is in fact what we found in our previous dark-energy related work, \cite{FE08.2}. However, with a little bit of attention, it turns out that this kind of reasoning is not correct. When the probability distribution for density fluctuations changes, the distribution of density peak heights is also changed, meaning that there are more (less) high peaks if $f_\mathrm{NL}$ is positive (negative). Hence, the different abundance of structures reflects the different distribution of peak heights, not the threshold for structure collapse. At the same time, for positive $f_\mathrm{NL}$ the density peaks themselves are also more clustered together with respect to the Gaussian case, implying a larger effective bias as found.

\begin{figure*}
	\includegraphics[width=0.45\hsize]{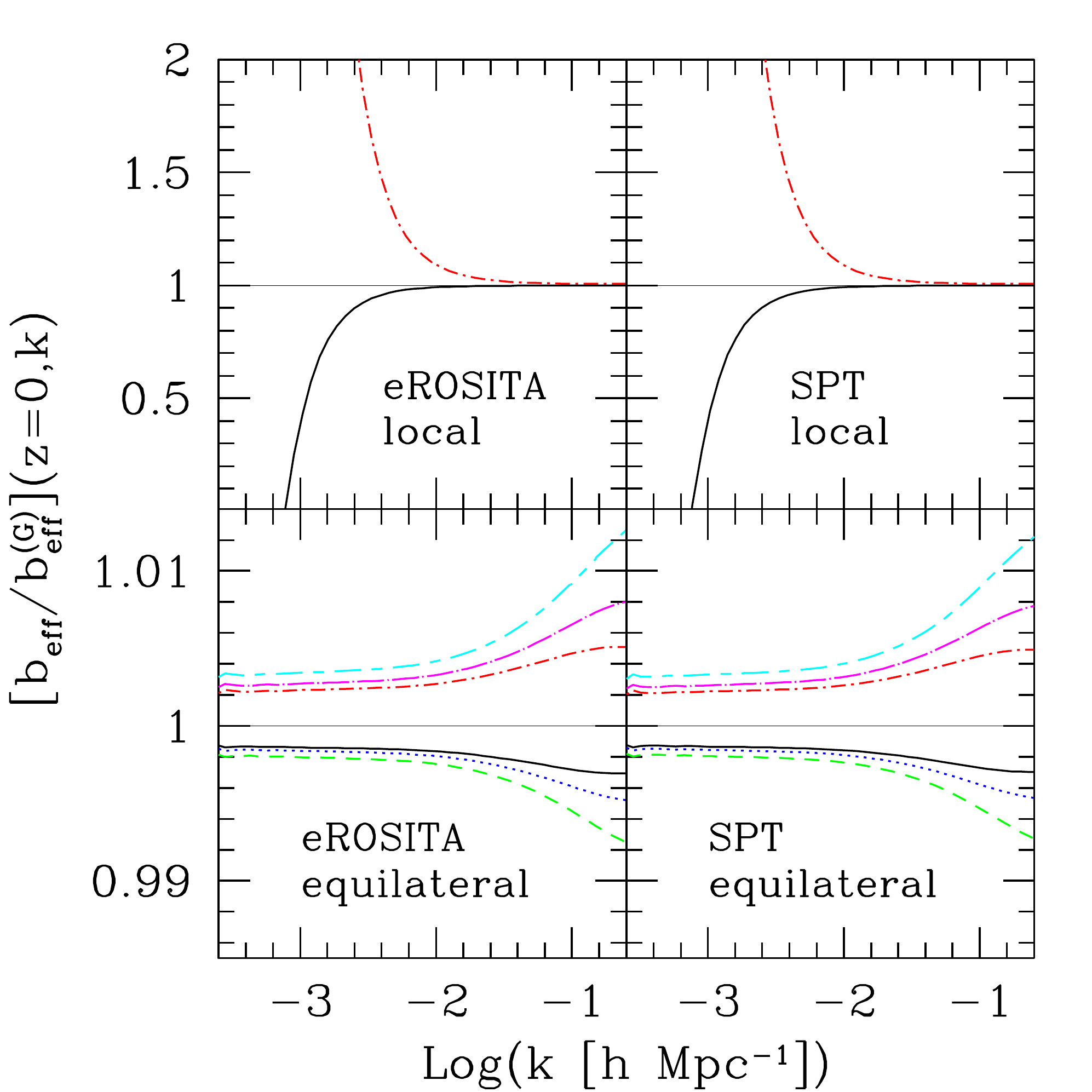}
	\includegraphics[width=0.45\hsize]{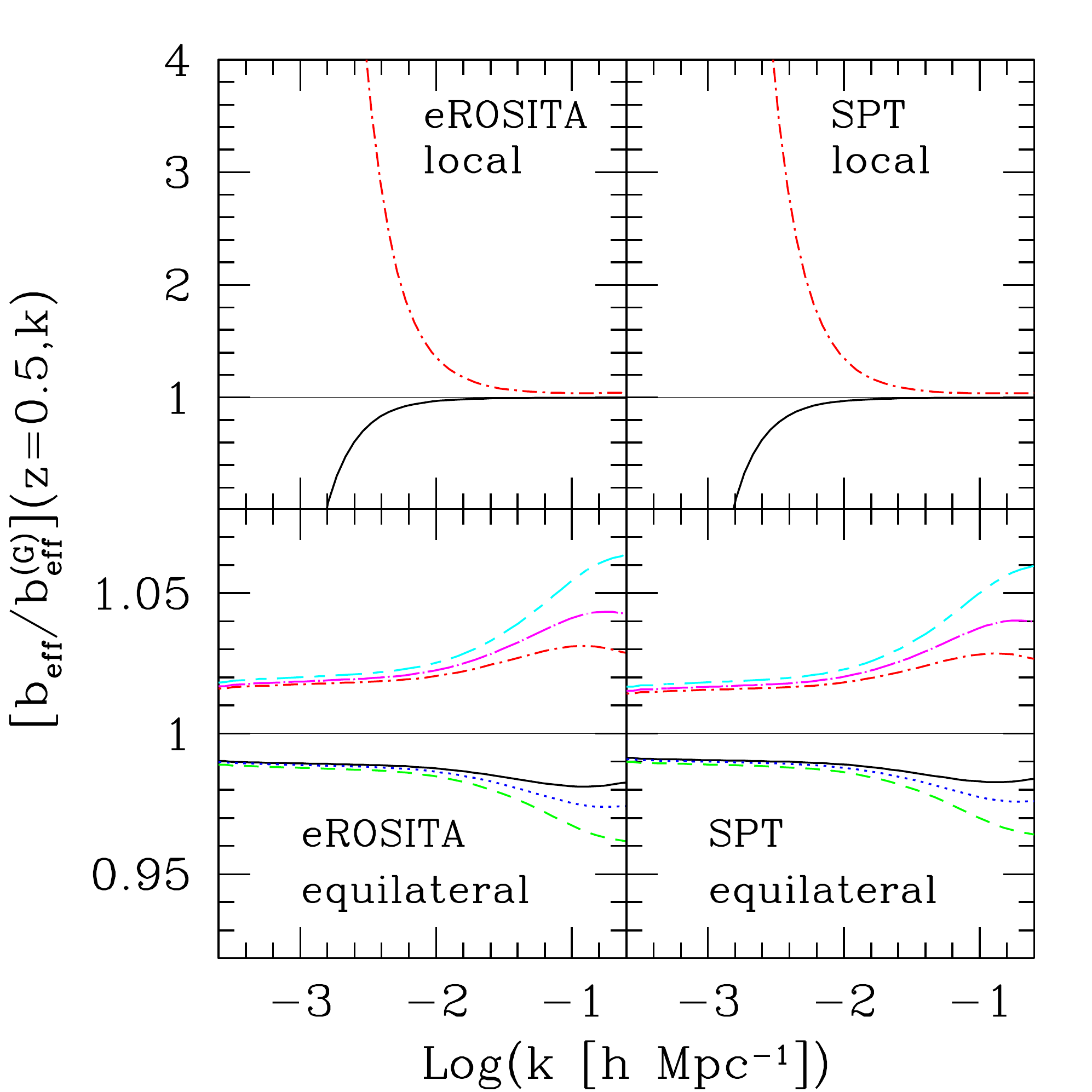}
	\includegraphics[width=0.45\hsize]{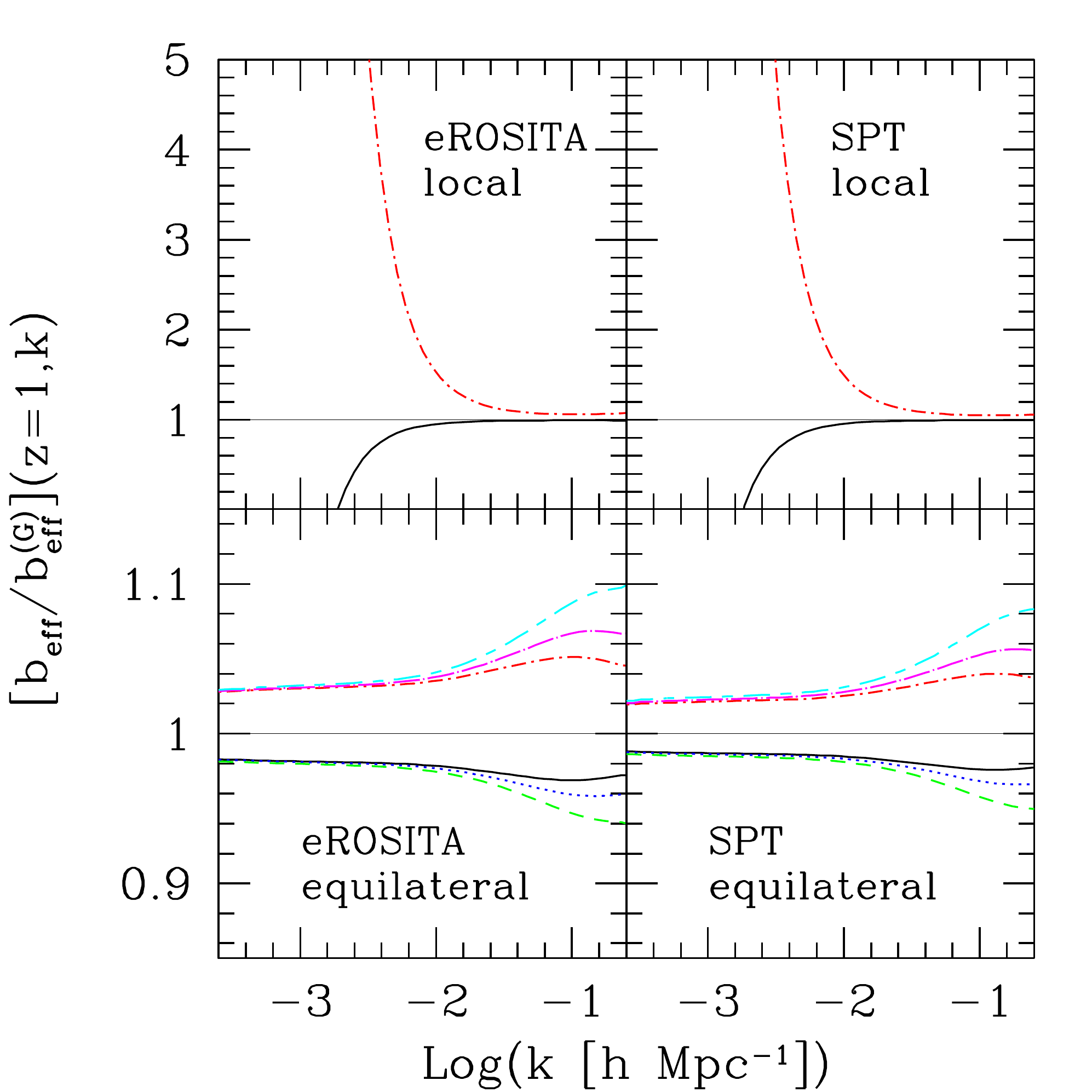}
	\includegraphics[width=0.45\hsize]{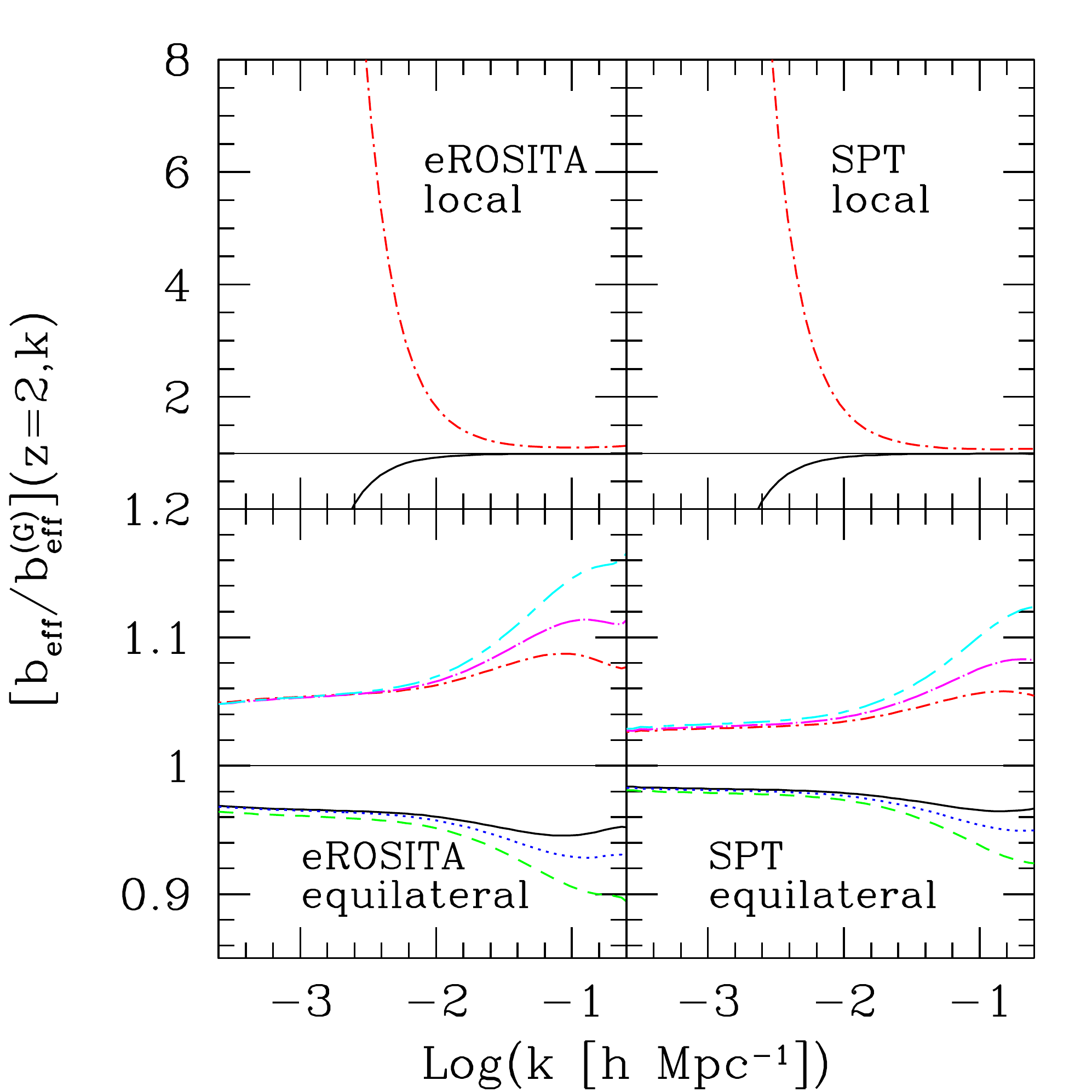}\hfill
	\caption{The effective bias of galaxy clusters included in the catalogues produced with the \emph{e}ROSITA and SPT surveys adopted in this work, relative to the scale-independent Gaussian effective bias. The thin black lines refer to the Gaussian model, while the other lines refer to initial conditions of non-Gaussian type, with colors and line style as in previous Figures. The redshift is fixed at $z = 0$ (top-left group of panels), $z = 0.5$ (top-right), $z = 1$ (bottom left) and $z = 2$ (bottom-right).}
\label{fig:bias_z}	
\end{figure*}

Similarly to what happens for the redshift distributions, the differences between Gaussian and non-Gaussian models are more evident in the \emph{e}ROSITA catalogue as compared to the SPT catalogue. This is a consequence of the fact that more massive objects are included in the former, that are more affected by non-Gaussian initial conditions. The increment due to non-Gaussianity can be up to $\sim 20\%$ for the effective bias. When the scale at which the effective bias is evaluated is increased, the difference with respect to the Gaussian scenario generally decreases for the equilateral shape, in agreement with the general behavior to the bias correction examined before. Likewise, we find a slight increase of the difference in the effective bias for the local models.

In Figure \ref{fig:bias_z} we show the ratio of the effective bias with respect to the Gaussian model for the two cluster catalogues at hand as a function of scale for different fixed redshifts, ranging from $z = 0$ to $z = 2$. 

\begin{figure*}
	\includegraphics[width=0.45\hsize]{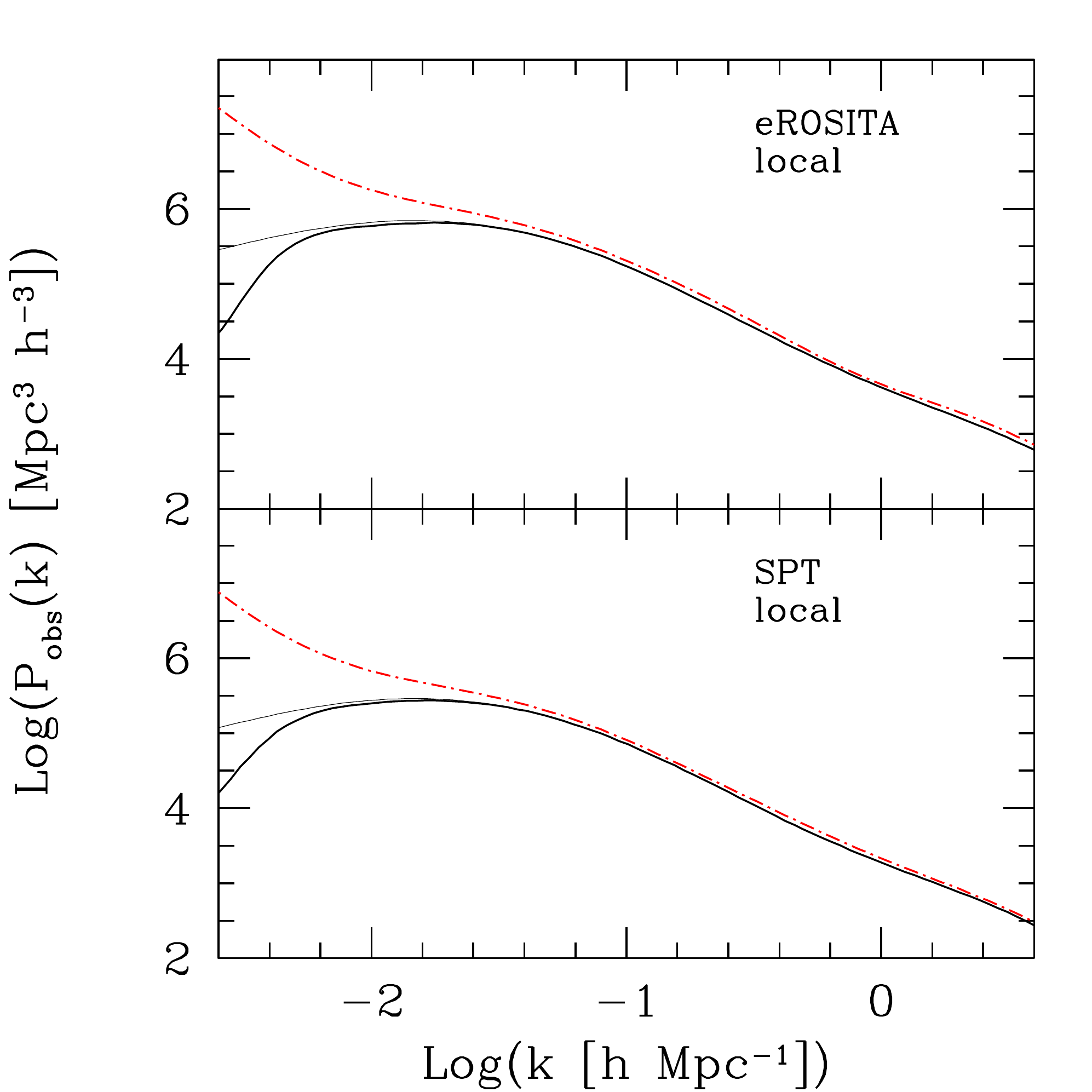}
	\includegraphics[width=0.45\hsize]{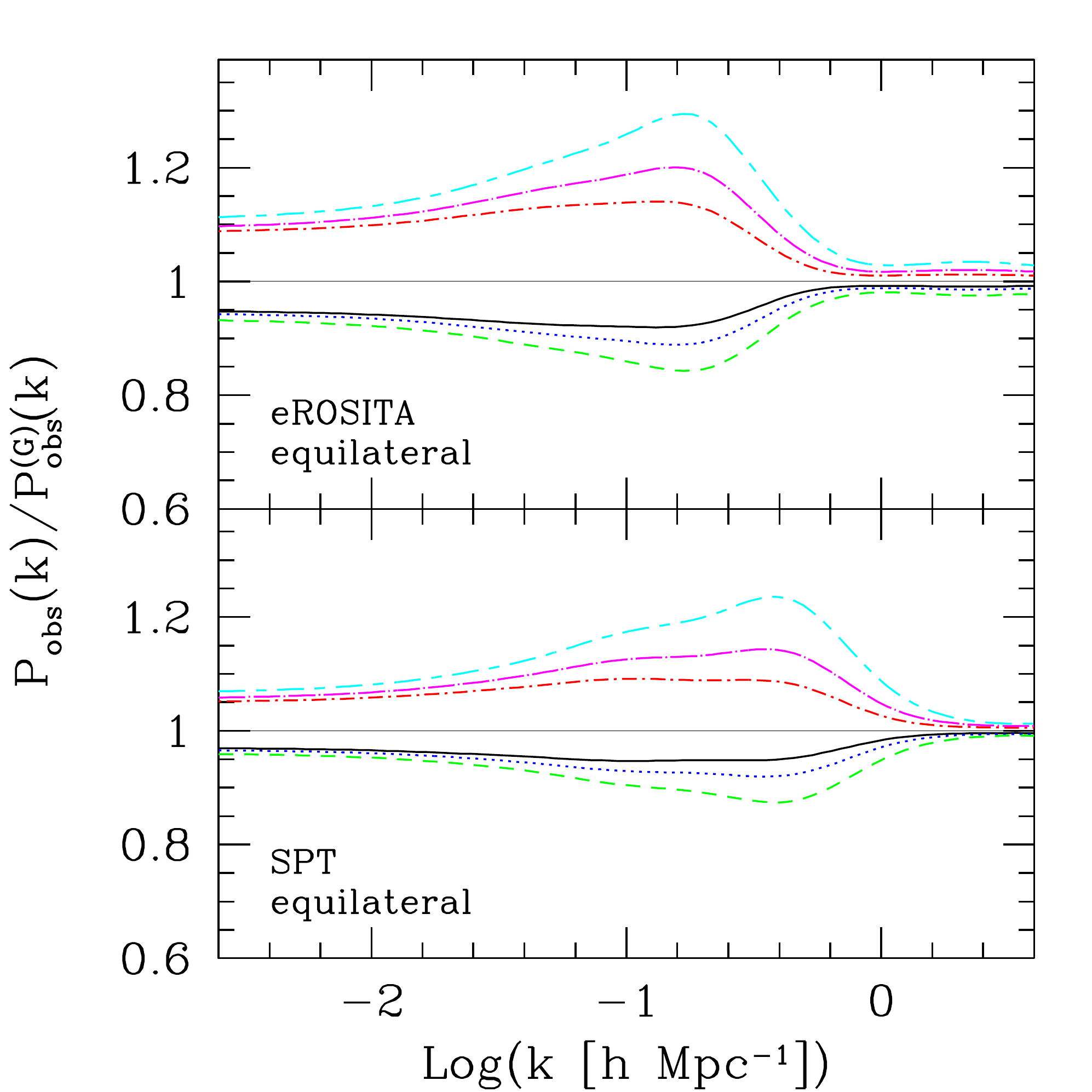}\hfill
	\caption{The observed power spectrum for galaxy clusters inside the \emph{e}ROSITA and SPT catalogues, as labelled in the plots. The two leftmost panels show results for non-Gaussian models with local shape, while the rightmost two show results for models with equilateral shape relative to the Gaussian case. In all panels, the thin black lines refer to the Gaussian model, while other line types and colors follow the same convention as in the previous Figures.}
\label{fig:obsp}	
\end{figure*}

Reflecting the trend that has already been noticed in the correction to the linear "monochromatic" bias, the ratio between the effective bias in a non-Gaussian model with local-shape bispectrum and the Gaussian one grows indefinitely at small values of $k$ for positive values of $f_\mathrm{NL}$, while it decreases below unity for negative $f_\mathrm{NL}$ at the same scales. For instance, at $z = 0$ and for $f_\mathrm{NL} = 145$, the non-Gaussian bias is already a factor of $2$ larger than the Gaussian one at scales $\gtrsim 400\:h^{-1}$ Mpc. At $z = 2$ the scale where this happens reduces to $\sim 100\:h^{-1}$ Mpc. The effect is milder for negative $f_\mathrm{NL}$ since in this case the value of the non-Gaussian amplitude is substantially closer to zero. Analogously, the ratio between non-Gaussian effective bias computed in models with bispectra of the equilateral shape and the Gaussian case have an opposite trend compared to the local shape, in that it increases with the wavenumber $k$. Also, it has a much milder variation with respect to the local case, being almost constant over the range of scales considered here. In the equilateral case, the deviation with respect to the Gaussian case is only of $\sim 1\%$ at $z = 0$, and grows up to $\sim 15\%$ at $z = 2$. We also note the usual difference between catalogues, with the \emph{e}ROSITA one displaying larger variations than the SPT one.

One can use the effective bias of galaxy clusters and the power spectrum of the dark matter density field to construct the power spectrum of clusters that should be observed in the different surveys. Following the notation of \cite{MA97.1} (see also \citealt{MO98.1,MO00.1,MO01.1,MO02.1}) we may write the approximate expression

\begin{equation}\label{eqn:obsp}
P_\mathrm{obs}(k) = \frac{1}{\Gamma} \int_{z_1}^{z_2} \frac{\mathcal{N}^2(z)}{G(z)}b^2_\mathrm{eff}(k,z) P(k,z) dz,
\end{equation}
where the normalization constant $\Gamma$ reads

\begin{equation}
\Gamma = \int_{z_1}^{z_2} \frac{\mathcal{N}^2(z)}{G(z)} dz.
\end{equation}
In the two previous equations $z_1$ and $z_2$ are the limiting redshifts of the cluster catalogue at hand. Practically, we shall have $z_1 \simeq 0$, while $z_2$ is the maximum redshift at which objects are present in the catalogue. We adopted the \cite{PE94.1} fit for computing the nonlinear matter power spectrum, as we believe it suffices to our purposes (see the discussion in \citealt{FE08.2}). Additionally, we decided to neglect the redshift-space distortion to be applied to the matter power spectrum \citep{KA87.1,ZA96.1,MA00.1}. As explained in \cite{MO00.1}, this correction results in a small change to the correlation function \citep{BO99.1}, corresponding to an at most $\sim 6\%$ increase on the observed correlation length for deep surveys. We safely ignore this correction because it is not very dependent on the non-Gaussian model, and we are mainly interested in relative differences.

In Figure \ref{fig:obsp} we show the results on the observed power spectrum for the galaxy clusters contained in both the \emph{e}ROSITA and SPT catalogues, as well as for all the non-Gaussian cosmologies considered in this work. In that Figure, for non-Gaussianity of the equilateral shape we only show the ratio of the observed power spectrum to the Gaussian case, in order to better highlight the differences that would hardly be visible otherwise. Let us first focus on the local shape model, that is perhaps the most interesting one. On small scales, the non-Gaussian power spectrum always approaches the Gaussian one. Perfect coincidence is never achieved, since the correction to the linear bias never vanishes \citep{MA08.1}, and even if it would, the redshift distributions would still be different. On large scales, the power spectrum for the non-Gaussian models with positive $f_\mathrm{NL}$ increases without bound, according to the behavior of the correction to the "monochromatic" bias discussed above. Similarly, when $f_\mathrm{NL}$ is negative the observed power spectrum decreases far below the Gaussian one.

In the non-Gaussian model with equilateral shape of the primordial bispectrum, the differences with respect to the Gaussian model are very small, such that they are almost not visible unless we take a zoom of some region or we perform the ratio to the Gaussian case itself. This is in agreement with the behavior of the redshift distribution and effective bias discussed above, and shows that these kinds of models should be more difficult to be distinguished by the Gaussian scenario using the spatial distribution of galaxy clusters. We additionally note that, coherently with the discussion presented in the previous sections, the effect of non-Gaussianity of equilateral shape on the observed cluster power spectrum is more marked for the \emph{e}ROSITA catalogue than for SPT.

\begin{figure*}
	\begin{center}
	\includegraphics[width=0.8\hsize]{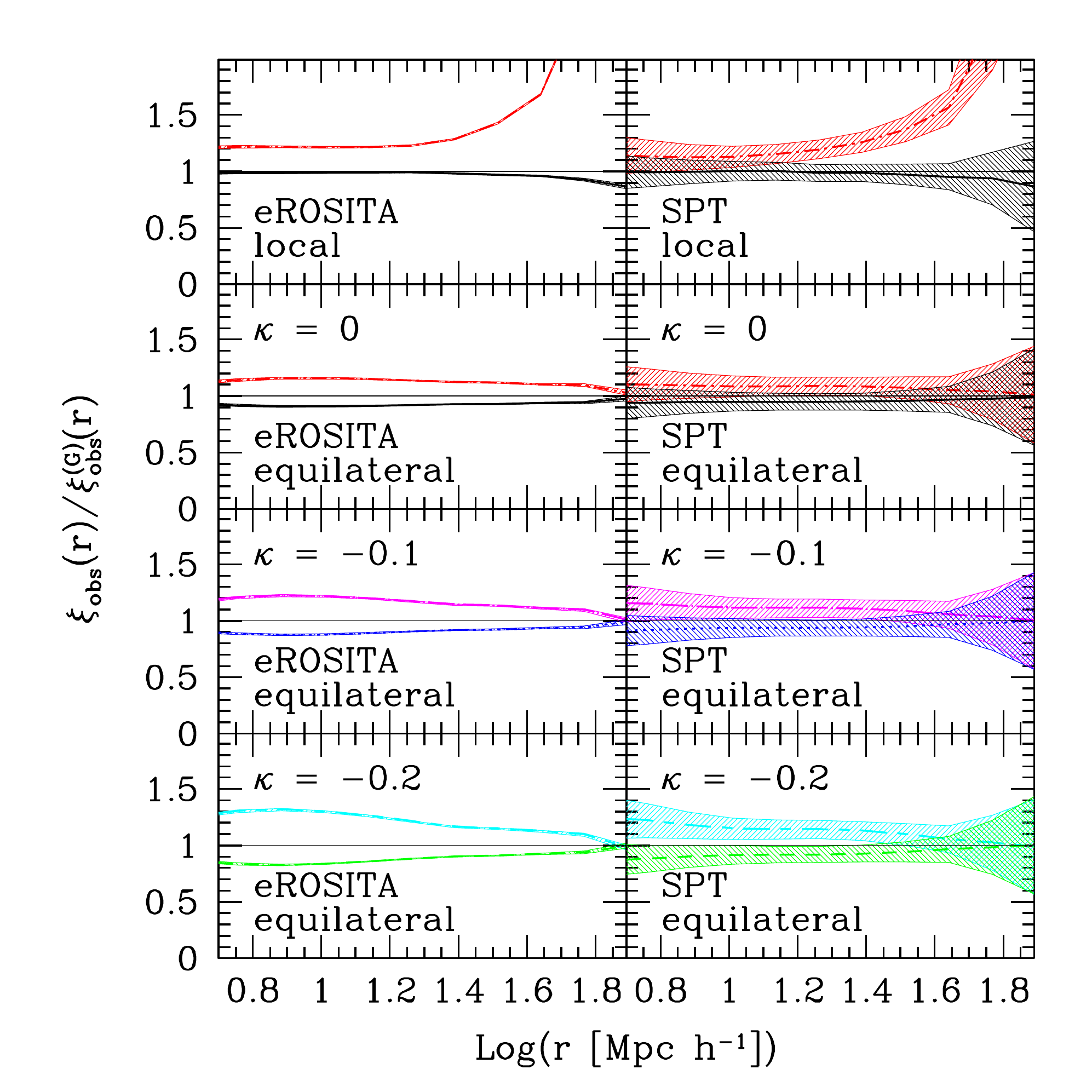}\hfill
	\caption{The ratio between the observed correlation functions for the various non-Gaussian models to the Gaussian case, represented by the thin black line, as a function of spatial separation. The top two panels refer to non-Gaussian models with local shape, while the other panels to models with equilateral shape, with different scale dependence of $f_\mathrm{NL}$, as labelled. The shaded regions denote errors computed according to the bootstrap method for each correlation function and then propagated to the ratios. Colors and line styles are the same as in previous Figures.}
\label{fig:correlation}	
	\end{center}
\end{figure*}

By performing the Fourier transform of Eq. (\ref{eqn:obsp}) with respect to wavenumber we then computed the observed correlation function of galaxy clusters, that we denote with $\xi_\mathrm{obs}(r)$. In Figure \ref{fig:correlation} we report the ratio of this function to the Gaussian case for the various non-Gaussian models considered in this work and the two cluster catalogues we adopted. Errors on the correlation functions are computed via the bootstrap method, and are then propagated to the ratios. The first thing to note in this Figure is that the relative errors for the \emph{e}ROSITA catalogue are extremely small, and much smaller than those for the SPT catalogue. This is in part due to the different area of the sky that is covered by the two surveys, with the \emph{e}ROSITA one being $\sim 5$ times larger than the SPT one. However, as noticed above, the SPT survey area might be underestimated here, and it is possible that eventually the errorbars for this catalogue will be smaller than depicted in Figure \ref{fig:correlation}.

\begin{figure*}
	\begin{center}
	\includegraphics[width=0.45\hsize]{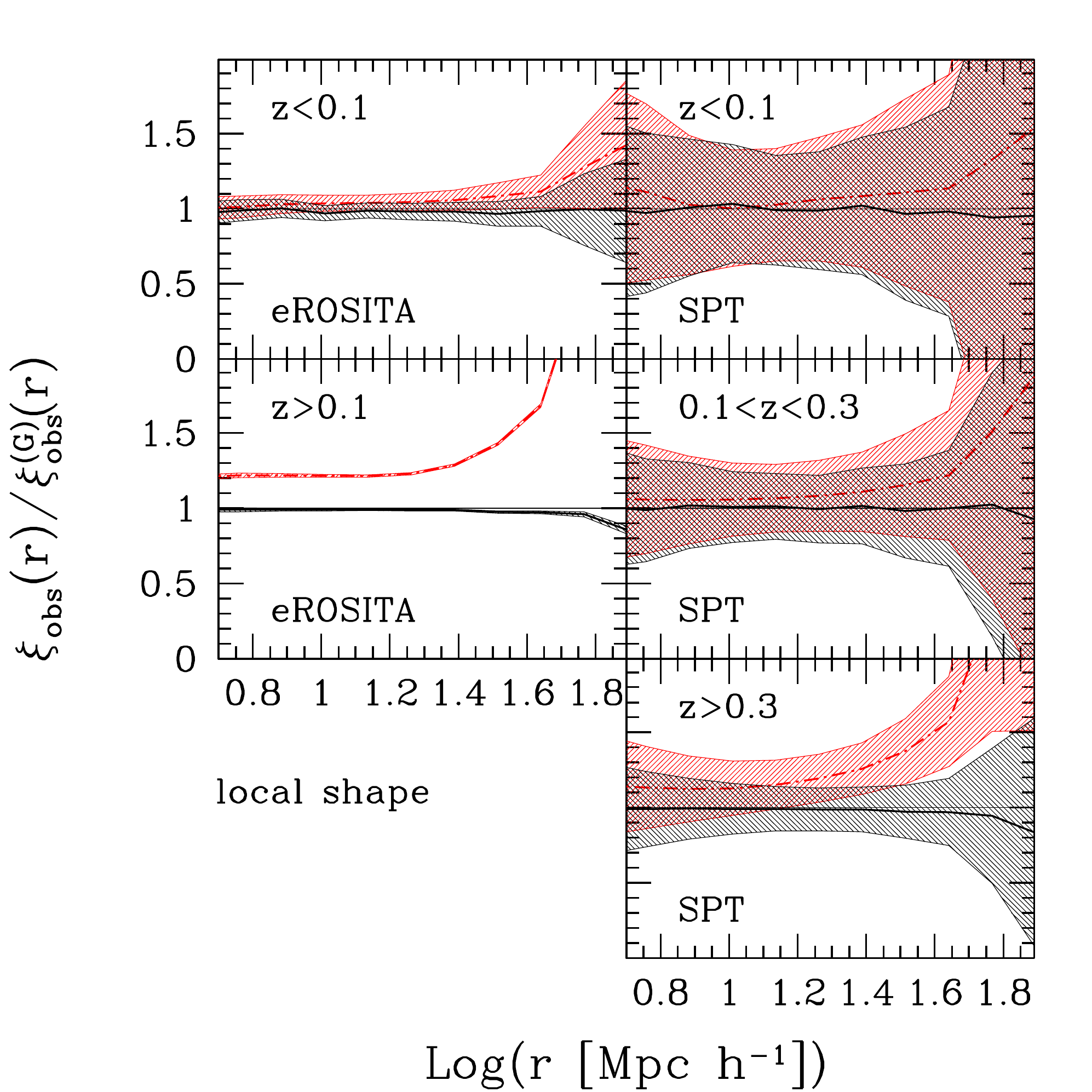}
	\includegraphics[width=0.45\hsize]{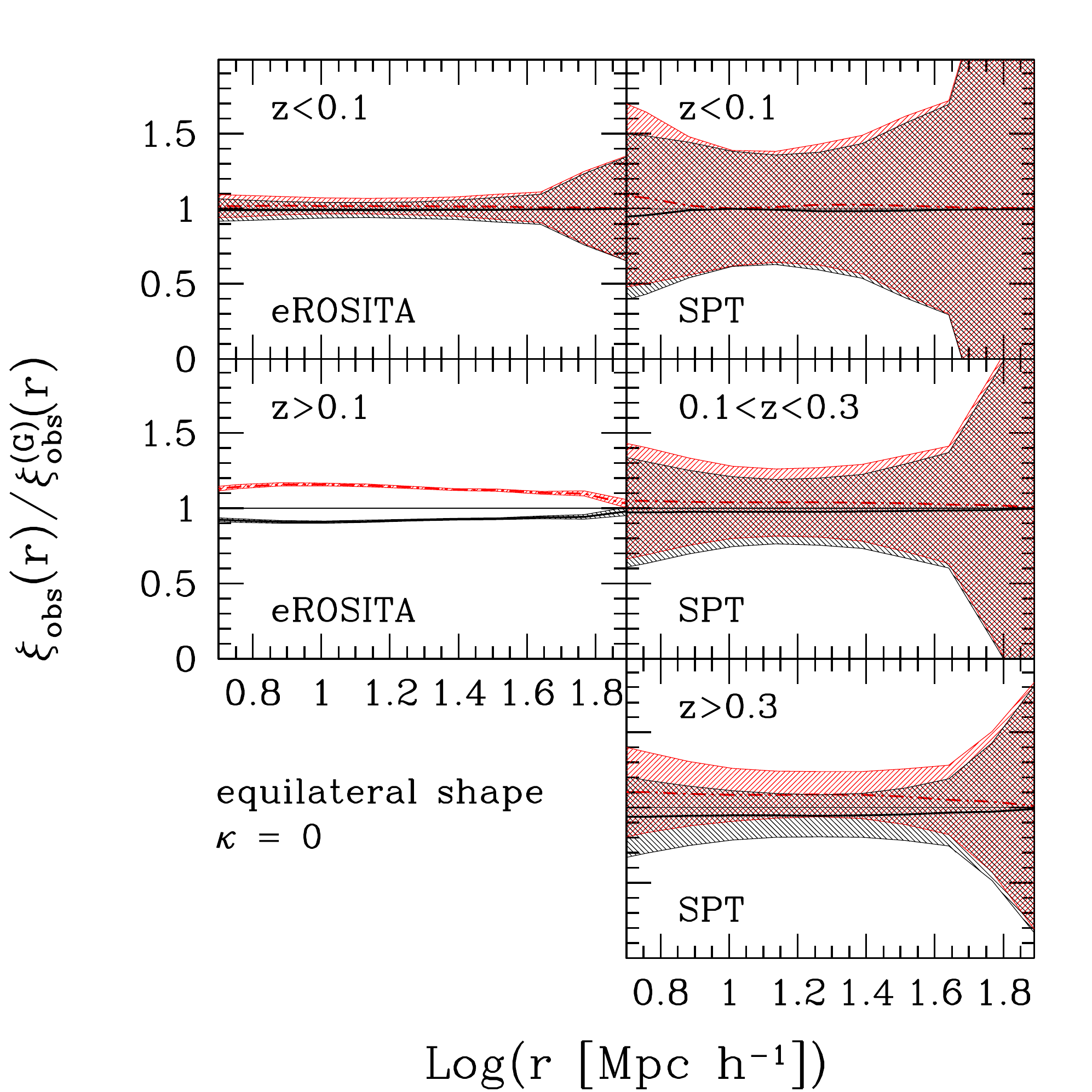}
	\includegraphics[width=0.45\hsize]{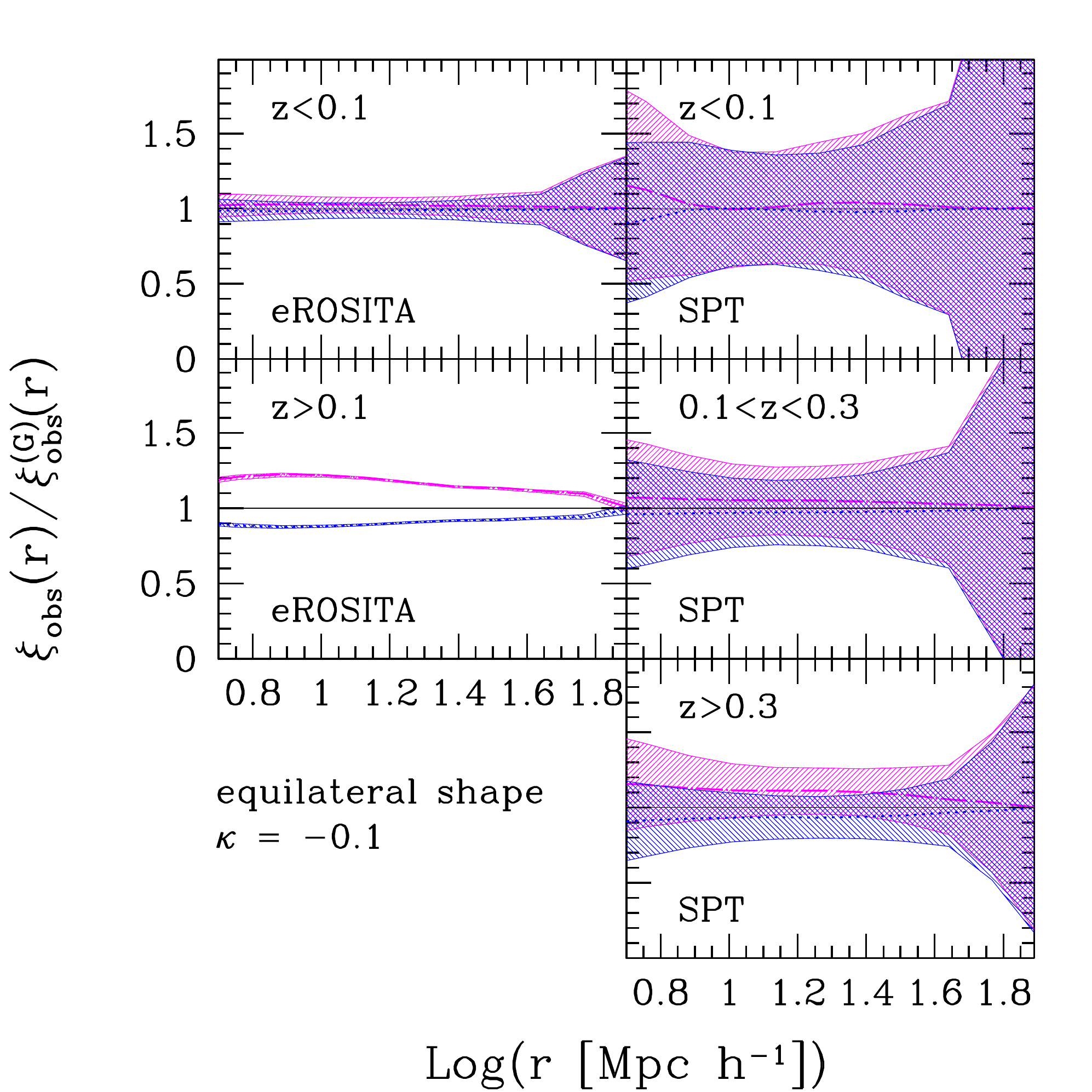}
	\includegraphics[width=0.45\hsize]{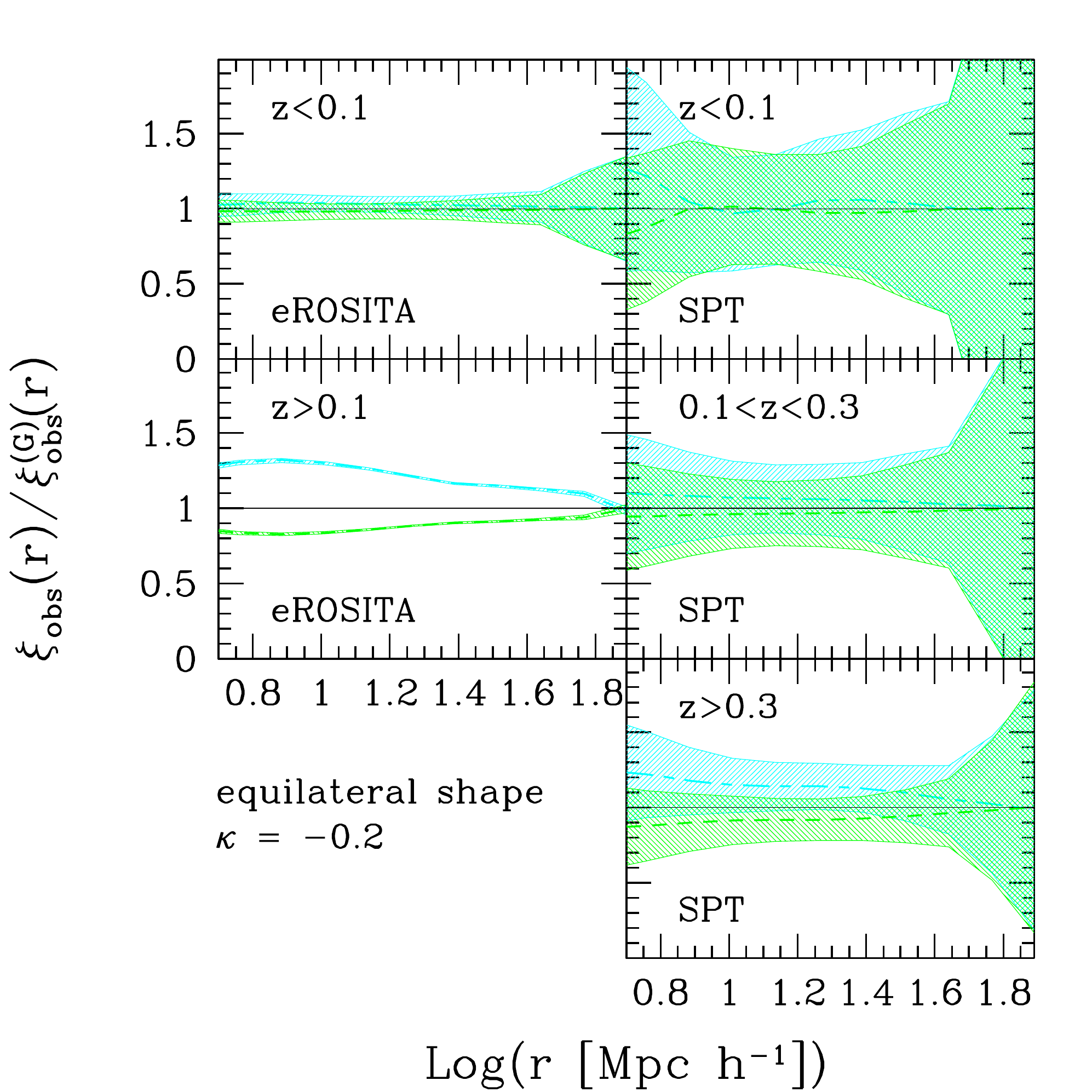}
	\caption{The ratio of the spatial correlation functions observed in the different catalogues to the Gaussian case. Only clusters included in the redshift bins labeled in the plots are considered in each panel. For \emph{e}ROSITA we consider only two redshift bins, while for SPT we consider three of them, as detailed in the text. The thin black line refers to the Gaussian initial conditions, while other line types and color codings are the same as in previous Figures. Errors are computed with the bootstrap method for each correlation function and then propagated to the ratios.}
\label{fig:correlation_bin}	
	\end{center}
\end{figure*}

For non-Gaussian models with primordial bispectrum of the local shape, the deviation of the observed correlation function with respect to the Gaussian case grows with increasing radius, a trend that reflects the one already observed in the power spectrum. With SPT, only the model with positive $f_\mathrm{NL}$ can be reliably distinguished by the Gaussian case, with the errorbars being too large to allow the same for the model with negative $f_\mathrm{NL}$. On the other hand, with \emph{e}ROSITA the errorbars are small enough to allow the separation also of the latter models, if sufficiently large scales are probed. Considering the equilateral shape instead, the differences between Gaussian and non-Gaussian models are much more reduced, and maximized at scales $\lesssim 10 h^{-1}$ Mpc. As a consequence, only in the \emph{e}ROSITA sample the models could be distinguished, while in the SPT one the errorbars would be too large. An exception to this is maybe given by the model with positive $f_{\mathrm{NL},0}$ and the most extreme scale dependence of $f_\mathrm{NL}$, namely $\kappa = -0.2$. In this case the deviation between models might just be large enough to be resolved.

A popular way to rapidly quantify the correlation strength is the correlation length $r_0$, defined such that $\xi_\mathrm{obs}(r_0) = 1$. The effect of non-Gaussianity with positive amplitude $f_\mathrm{NL}$ results in an increase in the measured correlation length of $\sim 20\%-30\%$ with respect to the Gaussian case, with the precise value depending on the model and on the catalogue considered.

It is interesting to note that the ratio of non-Gaussian correlation functions to the Gaussian one is almost constant for the equilateral shape and for the local shape at spatial separations $\lesssim 20 h^{-1}$ Mpc. At these scales, the difference with respect to the Gaussian case might be accounted for by a change in the normalization $\sigma_8$ of the primordial power spectrum, that enters quadratically in the normalization of the observed correlation function. Turning the argument around, this gives the precision with which is necessary to know $\sigma_8$ in order to disentangle the effect of non-Gaussianity. It turns out that for models with positive $f_\mathrm{NL}$ this precision is of the order of $\sim10\% - 15\%$, depending on the model, a precision that is already available. 

As a final step, we computed the observed power spectrum and spatial correlation functions when only clusters within selected redshift bins are considered for each catalogue. For the \emph{e}ROSITA survey we considered two bins, $z < 0.1$ and $z > 0.1$, while for the SPT survey we adopt the three bins, $z < 0.1$, $0.1 < z < 0.3$ and $z > 0.3$, thanks to its far wider redshift distribution of objects. As explained in \cite{FE08.2}, this choice insures an approximately equal number of pairs of objects in each bin. In Figure \ref{fig:correlation_bin} we show the ratio of the correlation functions obtained for the different kinds of non-Gaussianity assumed in the present work to the Gaussian case. Let us focus first on the \emph{e}ROSITA catalogue. In the low-$z$ bin we basically have no signal, since the difference between non-Gaussian and Gaussian models is very small and the errors are quite large. Instead, for the redshift bin $z > 0.1$ the deviations from the Gaussian case are large and the relative errors are small, allowing a significant separation. It should be noted that in this latter redshift bin, the absolute errors are actually slightly larger than in the former, however the correlation function is also larger, so that the relative error is effectively smaller. These conclusions apply to both shapes of the primordial non-Gaussian bispectrum.

Focusing on the SPT catalogue, we note that the deviations of non-Gaussian models from the Gaussian one increase with increasing redshift, and the size of the relative errors decreases accordingly. The non-Gaussian models with equilateral shape can never be distinguished from the Gaussian case, since the deviations therefrom are always too small compared to the errors. The situation is different for the local shape, where deviations from Gaussianity can be detected in the highest redshift bin ($z > 0.3$) and at sufficiently large spatial scales. The latter however only applies to the case with positive $f_\mathrm{NL}$, since the other one is still too similar to the Gaussian model.

\section{Summary and conclusions}\label{sct:con}

In this work we evaluated the main properties of galaxy cluster catalogues to be built with the two forthcoming survey performed with \emph{e}ROSITA and SPT, exploring cosmological models with various kinds of non-Gaussian initial conditions. In particular, we focused on the observable redshift distributions, on the effective bias and on the observed power spectrum of galaxy clusters obtained with the two catalogues in the different cosmologies. The two cluster catalogues are constructed adopting the predicted survey properties and simple yet realistic scaling relations between mass and X-ray/SZ observables. 

The non-Gaussian models adopted have both local and equilateral shape of the primordial bispectrum, with different amplitude of the non-Gaussian deviation, compatible with the bounds coming from CMB and other probes. The redshift distribution of objects in the two catalogues is only mildly affected by primordial non-Gaussianity, resulting only in at most $\sim 80\%$ modification for SPT and a factor of $\sim 2.5$ for \emph{e}ROSITA at the most extreme masses and redshifts. The reason for this difference in the two catalogues is that X-ray flux drops more steeply than SZ flux density with redshift, hence the latter catalogue is more dominated by high-mass objects compared to the former, which in turn are more affected by non-Gaussianity. 

The effective bias is affected in a way coherent with previous work. Namely, it displays a scale dependence that is absent in Gaussian models. The deviation of the effective bias with respect to the Gaussian case grows at large scales in models with local shape, while it mildly grows at intermediate-small scales for models with equilateral shape. As a consequence, while the deviations from the Gaussian case can be very large in models with local non-Gaussianity if the wavenumber is small enough, in case of non-Gaussianity with equilateral shape maximal deviations range from $\sim 1\%$ at $z = 0$ up to $\sim 15\%$ at high redshift. It is interesting to note that non-Gaussian models with a positive $f_\mathrm{NL}$ provide a larger abundance of massive structures \emph{and} a larger effective bias, meaning that not only in these models the large peaks that eventually collapse into bound structures are more numerous, but that also peaks themselves are more clustered together. The opposite obviously applies to the case of negative $f_\mathrm{NL}$. 

The power spectrum that is predicted to be observed with the use of the two cluster catalogues above reflects the behavior of the effective bias. As a matter of fact, for non-Gaussian models with equilateral shape, the power spectrum is very similar to the Gaussian one, with maximal deviations occurring at intermediate scales and reaching up to $\sim 20\%$ for the model with $\kappa = -0.2$. On the contrary, the cluster power spectrum deviates significantly from the Gaussian case for non-Gaussianity with local shape. In particular, when $f_\mathrm{NL}$ is positive, the power spectrum grows indefinitely at large scales, being already $\sim 2$ orders of magnitude larger than the Gaussian power spectrum at $k \sim3 \times 10^{-3} h$ Mpc$^{-1}$. For the same kind of models but negative $f_\mathrm{NL}$, the power spectrum decreases far below the Gaussian one. All of these conclusions apply quite independently of the catalogue adopted, except that in the \emph{e}ROSITA one the effect of non-Gaussianity tends to be slightly more marked than for the SPT catalogue, as a consequence of the different mass composition of the two, as explained above.

By computing the Fourier transform of the observed power spectrum we also evaluated the observed correlation function that is expected to be measured with the catalogues detailed above in the different non-Gaussian models. We estimated the expected errors on the observed correlation function by using the bootstrap method. Coherently with the behavior of the power spectra we find that the observed correlation functions for the local non-Gaussian models deviates strongly from the Gaussian model at large spatial separations. The deviation is expectedly more marked for the model with positive amplitude of the non-Gaussian contribution, since in that case $f_\mathrm{NL}$ is more distant from zero. For non-Gaussianity of equilateral shape instead, the deviations with respect to the Gaussian case stay always quite limited, and never grow above $\sim 20\%$. The relative errors on the observed correlation function for the \emph{e}ROSITA catalogue are much smaller than those for the SPT catalogue, mainly due to the largest area of the sky that the former cover. Therefore, not only the differences between models are slightly more enhanced in \emph{e}ROSITA compared to SPT, but also the errors are smaller in the former. This certainly makes \emph{e}ROSITA the ideal tool for this kind of study.

We also demonstrated that subdividing the two cluster catalogues in different redshift bins, the better results are always got when the highest bins are considered. This is consistent with deviations from Gaussianity being larger at higher redshift and higher masses (that are preferentially selected at high $z$). Even in this way however, only \emph{e}ROSITA seems to be able to detect deviations from Gaussian initial conditions. Separating a given cluster catalogue in different redshift bins does not give any particular advantage over considering the entire sample, however it demonstrates the importance of including high-$z$ objects, that produce the bulk of the signal.

Two possible distinctive signatures of non-Gaussian initial conditions that have not been discussed here are somewhat related to the present work. The first is the statistics of voids in the large scale structure. This issue has been recently addressed by \cite{KA09.1} (see also \citealt{GR08.2}), showing that the abundance of large empty regions can indeed be used to put constraints on the non-Gaussian amplitude $f_\mathrm{NL}$ at the level of few tens. The second is the use of maximum cluster mass as a function of redshift as a discriminator between models. While this is an interesting issue that deserves exploration, the comparison with real observations would be quite difficult, since objects at the extreme mass end are especially rare. Also, because of this paucity, the scaling relations at such high masses are not well defined.

Before concluding, it is worth mentioning that the errors estimated in this work do not take into account the presence of scatter around the scaling relations used to link the dark-matter halo mass with cluster observables. This scatter will have the effect of somewhat increase the size of errorbars, so that our reported values are likely to be lower limits. Still, in order to produce large deviations from our results, the scatter would need to be extremely skewed, and more complete datasets would be needed in order to understand whether this is indeed the case.

A complete statistical analysis of the predictive power of forthcoming cluster surveys in terms of shape of the primordial bispectrum and level of non-Gaussianity is certainly a step to perform, but goes beyond the main purpose of this work. We can conclude that the effect of primordial non-Gaussianity on the clustering properties of galaxy clusters is generically mild, but depends strongly on the shape of the primordial bispectrum that is chosen. It is likely that constraints competitive with those from the CMB can be given in this way using the \emph{e}ROSITA catalogue. For the SPT catalogue this is probably not possible, unless the survey area is increased such that the error on the observed spectrum can be reduced below $\sim 10\%$, or clustering measurements can be pushed out to very large scales, but in this case, only if the non-Gaussianity is effectively of the local shape.

\section*{Acknowledgments}
{\small We acknowledge partial support by ASI contracts I/016/07/0 ÒCOFISÓ, ÒEuclid-DuneÓ I/064/08/0Ó, ASI-INAF I/023/05/0, ASI-INAF I/088/06/0 and ASI contract Planck LFI Activity of Phase E2. We are grateful to D. Crociani for kindly providing the codes computing non-Gaussian mass functions. We wish to thank the anonymous referee for useful remarks that allowed us to improve the presentation of our work.}

{\small
\bibliographystyle{aa}
\bibliography{./master}
}

\end{document}